\documentclass[a4paper,11pt]{article}
\pdfoutput=1 % ensure pdflatex output for arXiv

\usepackage{jheppub} % JHEP package

\usepackage[T1]{fontenc} % better font encoding

\usepackage{graphicx}
\usepackage{amsmath,amssymb}
\usepackage{commath}
\DeclareMathOperator{\arcsinh}{arcsinh}
\usepackage{array}
\usepackage{mathrsfs}
\usepackage{slashed}
\usepackage{xcolor} % for custom text color
\def\ba{\begin{eqnarray}}
\def\ea{\end{eqnarray}}

\def\be{\begin{equation}}
\def\ee{\end{equation}}

\newcommand{\checked}[1]{}

\newcommand{\la}{\label}

\usepackage{lineno}
\title{Bottomonium suppression and elliptic flow in an anisotropic quark-gluon plasma using the quantum trajectories method}

\author{Ajaharul Islam}
\affiliation{Institute of Particle Physics and Key Laboratory of Quark and Lepton Physics (MOE), Central China Normal University, Wuhan, 430079, China}
\emailAdd{aislam2@kent.edu}

\abstract{
We study bottomonium dynamics in a momentum-space anisotropic quark-gluon plasma (QGP) using the quantum trajectories (QTraj) framework. The real part of the heavy-quark potential is obtained from a minimal extension of the Karsch-Mehr-Satz (KMS) potential, while the angle-averaged imaginary part is derived to leading order in the anisotropy parameter $\xi$ and modeled to interpolate smoothly between the small- and large -$\xi$ regimes. The resulting anisotropic complex potential is used to solve the real-time Schrödinger equation using QTraj for the evolution of bottomonium in heavy-ion collisions. Nuclear modification factors $R_{AA}$, double ratios, and elliptic flow coefficients $v_2$ for the $\Upsilon(1S)$, $\Upsilon(2S)$, and $\Upsilon(3S)$ states are computed, including feed-down contributions,  in Pb-Pb collisions at $\sqrt{s_{NN}} = 5.02 \, \text{TeV}$. The QTraj-Aniso predictions successfully reproduce the observed sequential suppression pattern and non-zero elliptic flow, while for $R_{AA}$ and double ratios, the anisotropic implementation provides a modest but systematic improvement over the isotropic baseline, moving the predictions closer to experimental measurements from the ALICE, ATLAS, and CMS collaborations and further demonstrating the relevance of path-length dependent suppression and medium anisotropy in quarkonium phenomenology.
}


\begin{document}
%%%%%%%%%%%%%%%%%%%%%%%%%%%%%%%%%%%%%%%%%%%%%%%%%%
\setcounter{tocdepth}{2} % show subsections in TOC
\maketitle
\flushbottom

%%%%%%%%%%%%%%%%%%%%%%%%%%%%%%%%%%%%%%%%%%%%%%%%%%%%%%%%%%%%%%
\section{Introduction}
\label{intro}
%%%%%%%%%%%%%%%%%%%%%%%%%%%%%%%%%%%%%%%%%%%%%%%%%%%%%%%%%%%%%%
The primary goal of relativistic heavy-ion collisions at facilities like the Relativistic Heavy Ion Collider (RHIC) and the Large Hadron Collider (LHC) is to create and characterize the quark-gluon plasma (QGP), a state of deconfined quarks and gluons that existed in the early universe microseconds after the Big Bang \cite{STAR:2005gfr, ALICE:2010suc}. Among the most valuable probes of this exotic matter are heavy quarkonia, such as the bottomonium family ($\Upsilon(1S,2S,3S)$). Due to their large masses, these states are produced predominantly during the initial hard partonic scatterings, and their subsequent evolution through the QGP provides a unique spacetime tomography of the medium's properties \cite{Matsui:1986dk, STAR:2005gfr, Prino:2016cni}.

The original proposal by Matsui and Satz \cite{Matsui:1986dk} suggested that color screening in the deconfined medium would suppress quarkonium binding, serving as a direct signature of deconfinement. Measurements at RHIC and the LHC have since revealed a sequential suppression pattern that depends on the states' binding energies \cite{PHENIX:2006gsi, ALICE:2014wnc, CMS:2018zza, CMS:2019uhm}. Bottomonium suppression is considered a particularly clean probe compared to charmonium due to the negligible contribution from regeneration processes at LHC energies \cite{Prino:2016cni, Du:2015wha, Andronic:2015wma}.

A successful description of quarkonium dynamics requires incorporating both the screening of the $Q\bar{Q}$ potential and in-medium dissociation effects. The in-medium potential becomes complex-valued, $V(r,T) = V_R(r,T) + i V_I(r,T)$, where the real part $V_R$ governs the binding energy, and the imaginary part $V_I$ encodes thermal broadening and decoherence through Landau damping and singlet-to-octet transitions \cite{Brambilla:2011sg, Burnier:2014ssa}. Effective field theory approaches \cite{Brambilla:2004jw, Brambilla:2008cx, Rothkopf:2019ipj} and lattice QCD studies \cite{Burnier:2014ssa, Datta:2003ww, Aarts:2011sm} have provided valuable insight into the real and imaginary parts of the heavy-quark potential, with the latter encoding thermal broadening and decoherence. In contrast to the adiabatic treatment, some studies adopt a non-adiabatic approach~\cite{Boyd:2019arx, Bagchi:2023vfv}~and develop a framework for the real-time quantum evolution of quarkonium based on a complex in-medium potential~\cite{Islam:2020gdv, Islam:2020bnp}. Beyond static potential models, an essential challenge is the real-time evolution of quarkonia as open quantum systems. Stochastic methods, including the Lindblad equation in open quantum system~\cite{Brambilla:2017zei, Akamatsu:2014qsa, Yao:2021lus, Hitschfeld:2024gtt, Yang:2024ejk} and the quantum trajectories (QTraj) approach \cite{Omar:2021kra}, provide a framework to describe decoherence and suppression dynamically. Many recent studies solve the Lindblad equation within the potential non-relativistic quantum chromodynamics (pNRQCD) and open quantum system (OQS) framework using the QTraj approach~\cite{Brambilla:2022ynh, Brambilla:2023hkw, Strickland:2024oat, Brambilla:2024tqg, Brambilla:2025sis, Thapa:2025jua}.

A widely used parametrization is the Karsch–Mehr–Satz (KMS) potential \cite{Karsch:1987pv}, often employed as the baseline for isotropic QGP studies. However, the QGP produced in heavy-ion collisions is far from isotropic in the early stages. Strong longitudinal expansion induces significant momentum-space anisotropies, parameterized by $\xi$, which can markedly modify the plasma's properties, including the screening of charges and the heavy-quark potential \cite{Romatschke:2003ms, Romatschke:2004jh, Dumitru:2007hy, Majumder:2007zh, Nopoush:2017zbu}. Extensions of the potential framework to anisotropic plasmas have therefore been developed to account for such effects \cite{Dumitru:2009ni, Laine:2006ns, Laine:2007gj, Krouppa:2016jcl, Debnath:2025cmd}. To reduce computational cost and technical complexity, some studies replace the fully three-dimensional anisotropic potential with an effective one-dimensional potential by introducing an angle-averaged screening mass~\cite{Dong:2021gnb, Dong:2022mbo, Islam:2022qmj, Carrington:2024rxw}.

In this work, we present a comprehensive framework to investigate bottomonium suppression in an anisotropic QGP. We construct a new complex potential model that generalizes the isotropic KMS potential to an anisotropic medium. For the real part, $V_R^{\rm aniso}(r, \xi)$, we employ an extension incorporating an anisotropic screening mass \cite{Dumitru:2009ni}. For the imaginary part, $V_I^{\rm aniso}(r, \xi)$, we derive an analytic, angle-averaged expression that correctly captures the modification in both the small-$\xi$ limit, $\sim (1 - \xi/6)$, and the large-$\xi$ asymptotic behavior, $\sim 1/\sqrt{\xi}$ \cite{Dumitru:2009fy, Nopoush:2017zbu}. The resulting potential model was first developed in Ref. \cite{Islam:2024ptl}, and in what follows we shall refer to it as the anisotropic (Aniso) model.

To simulate the real-time evolution of bottomonium wave functions under this potential, we employ the quantum trajectories (QTraj) method \cite{Blaizot:2018oev, Brambilla:2021wkt}, which provides a powerful framework for describing the system's decoherence and dissociation dynamically. We compute the nuclear modification factors ($R_{AA}$), double ratios, and elliptic flow coefficients ($v_2$) for $\Upsilon(1S)$, $\Upsilon(2S)$, and $\Upsilon(3S)$ in Pb–Pb collisions at $\sqrt{s_{NN}}=5.02$ TeV, including feed-down effects. Our results show good agreement with data from ALICE, ATLAS, and CMS, capturing both the sequential suppression hierarchy and the non-trivial elliptic flow. These findings highlight the critical importance of path-length dependent suppression and momentum-space anisotropy in shaping bottomonium observables, providing new insight into the transport properties of heavy quarkonia in the evolving QGP.

This paper is organized as follows: In Section~\ref{sec02}, we present the isotropic KMS potential. In Section~\ref{sec03}, we detail the formulation of the real part of the anisotropic potential. Section~\ref{sec04} is devoted to the derivation of the imaginary part in the small- and large-$\xi$ limits and the construction of our full model. The complete anisotropic complex potential and its matching to the vacuum limit are presented in Section~\ref{sec05}. Our numerical method for solving the Schr\"{o}dinger equation is outlined in Section~\ref{sec06}, and the procedure for computing $R_{AA}$ with feed-down is described in Section~\ref{sec07}. We present our results for $R_{AA}$ and $v_2$ in Section~\ref{sec08} and give our conclusions and outlook in Section~\ref{sec09}.
%%%%%%%%%%%%%%%%%%%%%%%%%%%%%%%%%%%%%%%%%%%%%%%%%%%%%%%%%%%%%%%%%%%%%%%%%%%%%%%%%%%%%%
\section{Isotropic KMS potential}
\label{sec02}
%%%%%%%%%%%%%%%%%%%%%%%%%%%%%%%%%%%%%%%%%%%%%%%%%%%%%%%%%%%%%%%%%%%%%%%
In vacuum we take the heavy-quark potential to be given by a Cornell potential with a finite string breaking distance \cite{Islam:2020gdv, Islam:2020bnp}
\be
V_{\rm vac}(r) =
\begin{cases}  
	-\frac{a}{r} + \sigma r &\mbox{if } r \leq r_{\rm SB} \\
	-\frac{a}{r_{\rm SB} } + \sigma r_{\rm SB}   & \mbox{if } r > r_{\rm SB}
\end{cases} \, ,
\label{eq:vvac}
\ee
where \mbox{$a = 0.409$} is the effective coupling, \mbox{$\sigma = 0.21$~GeV$^2$} is the string tension, and \mbox{$r_{\rm SB}  =$ 1.25 fm} is the string breaking distance.  Using this set of vacuum parameters and assuming \mbox{$M_b = 4.7$ GeV} one obtains vacuum masses of \mbox{$\{9.46,10.0,9.88,10.36,10.25,10.13\}$ GeV} for $\Upsilon(1s)$, $\Upsilon(2s)$, $\chi_b(1p)$, $\Upsilon(3s)$, and $\chi_b(2p)$, respectively. We have the internal energy based real part of the isotropic Karsch-Mehr-Satz (KMS) potential~\cite{Matsui:1986dk, Karsch:1987pv, Shuryak:2004tx, Strickland:2011aa}
%%%
\begin{equation}
V_{\rm{KMS}}^{\rm{iso}}(r)=-\frac{a}{r}(1+m_D~ r)e^{-m_D~ r}+\frac{2\sigma}{m_D}\left[1-e^{-m_D~ r}\right]-\sigma r e^{-m_D~ r} \, .
\label{isorKMS}
\end{equation}
%%%
where the usual in-medium gluonic isotropic Debye mass is given by  
\begin{equation}
m_D(T)=F_{m_D}T\sqrt{N_c\left(1+\frac{N_F}{6}\right)\pi\alpha}~,~~\alpha = \frac{4}{3}\alpha_s (2\pi T)~,
\label{isomd}
\end{equation}
where $N_C = 3,~ N_F = 2,~C_F = 4/3$, and $F_{m_D}$ is an adjustable factor called $m_D$-factor. Here, we consider $F_{m_D} = 1$. In Eq.~\eqref{isorKMS}, we take \mbox{$a = \alpha= 0.409$}.

To match smoothly onto the zero temperature limit, we use
\begin{equation}
V_R^{\rm{iso}}(r) =
\begin{cases}  V_{\rm KMS}^{\rm{iso}}(r)  &\mbox{if } V_{\rm KMS}^{\rm{iso}}(r)  \leq V_{\rm vac}(r_{\rm SB}) \\
	V_{\rm vac}(r_{\rm SB}) & \mbox{if } V_{\rm KMS}^{\rm{iso}}(r) > V_{\rm vac}(r_{\rm SB})
\end{cases} \, .
\label{eq:isovmedre}
\end{equation}
In the limit $T\rightarrow0$, Eq.~\eqref{eq:isovmedre} reduces to Eq.~\eqref{eq:vvac}. 

For the imaginary part of the potential, we take the result of the leading-order resummed perturbative QCD calculation of Laine et. al~\cite{Laine:2006ns}
\begin{equation}
V_I^{\rm{iso}}(r) = - C_F \alpha_s T \phi(m_D~ r) \, ,
\label{eq:isovmedim}
\end{equation}
where
\begin{equation}
   \phi(\hat{r}) =  1 - 2 \int_0^\infty dz \frac{\sin(z)}{(z^2 + \hat{r}^2)^2}~,
\end{equation}
with $\hat{r} = r~m_D$.  We evaluate the strong coupling $\alpha_s$ at the scale $\mu = 2 \pi T$ and use three-loop running \cite{pdg} with \mbox{$\Lambda_{\overline{MS}} = 344$ MeV}.  This value of $\Lambda_{\overline{MS}}$ is chosen in order to reproduce the lattice result for the running coupling $\alpha_s(5\text{ GeV}) = 0.2034$~\cite{McNeile:2010ji}.

The resulting total complex-valued potential is of the form 
\begin{equation}
V^{\rm{iso}}(r) = V_R^{\rm{iso}}(r) + i V_I^{\rm{iso}}(r) \, .
\label{eq:isovform}
\end{equation}
After solving the resulting evolution using the QTraj framework, we refer to the corresponding results as ``QTraj-Iso'' throughout this work.
%%%%%%%%%%%%%%%%%%%%%%%%%%%%%%%%%%%%%%%%%%%%%%%%%%%%%%%%%%%%%%
\section{Real part of the anisotropic KMS potential}
\label{sec03}
%%%%%%%%%%%%%%%%%%%%%%%%%%%%%%%%%%%%%%%%%%%%%%%%%%%%%%%%%%%%%%
In an anisotropic QGP, the distribution function is taken to be of Romatschke-Strickland (RS) form \cite{Romatschke:2003ms, Romatschke:2004jh}
\be
f_{\rm aniso}({\bf p})\equiv f_{\rm iso}\!\left(\frac{1}{\lambda}\sqrt{{\bf p}^2+\xi ({\bf p}\cdot {\bf n})^2}\right) ,
\ee
\checked{m}
where $f_{\rm iso}$ is an isotropic distribution which in thermal equilibrium is given by a Bose-Einstein distribution, $\lambda$ is a temperature-like scale which corresponds to temperature in the equilibrium limit, and the parameter $\xi$ quantifies the degree of momentum-space anisotropy,
\begin{equation}
\xi = \frac{1}{2}\frac{\langle \bf p^{2}_{\perp}\rangle}{\langle p^2_z\rangle}-1~,
\end{equation}
where $p_z \equiv \bf p.n$ and $\bf p_{\perp}\equiv \bf p-\bf n(p.n)$ denote the particle momenta along and perpendicular to the direction $\bf n$ of anisotropy\cite{Romatschke:2003ms}, respectively.

After considering the ``minimal'' extension \cite{Dumitru:2009ni} of the internal energy based isotropic KMS potential \cite{Karsch:1987pv, Strickland:2011aa}, we write down the real part of the anisotropic KMS potential
\begin{equation}
V^{\mathrm{aniso}}_{\mathrm{KMS}}(r)=-\frac{a}{r}(1+\mu r)e^{-\mu r}+\frac{2\sigma}{\mu}\left[1-e^{-\mu r}\right]-\sigma r e^{-\mu r} ~,
\end{equation}
where \mbox{$a = 0.409$} is the effective coupling, \mbox{$\sigma = 0.21$~GeV$^2$} is the string tension, and $\mu$ is the anisotropic screening scale (anisotropic Debye mass) defined via
\begin{equation}
\mu^2(\xi, T) = f(\xi) m^2_D(T) ~.\label{mumd}
\end{equation}
Here, the isotropic Debye mass, $m_D(T)$, is defined by Eq.~(\ref{isomd}).\\

The function $f(\xi)$ in Eq.(\ref{mumd}) is given by \cite{Strickland:2018ayk} (also see in Appendix-\ref{app:a})
\begin{equation}
f(\xi)=\frac{\sqrt{2}(1+\xi)\arcsin{\left(\sqrt{\frac{\xi}{1+\xi}}\right)}}{\sqrt{\xi}\sqrt{1+\xi+\frac{(1+\xi)^2 \arcsin{\left(\sqrt{\frac{\xi}{1+\xi}}\right)}}{\sqrt{\xi}}}}~,\label{ffunc}
\end{equation}
which has the following properties
\begin{equation}
\lim_{\xi \rightarrow 0}f(\xi) = 1-\frac{2}{45}\xi^2 + \frac{44}{945}\xi^3 - \frac{8}{189}\xi^4 + {\cal O}\left(\xi\right)^{9/2}~,
\end{equation}
and
\begin{equation}
\lim_{\xi \rightarrow \infty}f(\xi) = \sqrt{\pi}\left(\frac{1}{\xi}\right)^{1/4} - \frac{2}{\sqrt{\pi}}\left(\frac{1}{\xi}\right)^{3/4} + \frac{4}{3\sqrt{\pi}}\left(\frac{1}{\xi}\right)^{7/4} + {\cal O}\left(\frac{1}{\xi}\right)^{9/4}~,
\end{equation}
in the small and large-$\xi$ limit respectively.

When $\xi$ is negative, we use 
\begin{equation}
f(\xi) = \frac{\sqrt{2}\arcsinh{\left(\sqrt{\frac{\abs{\xi}}{1-\abs{\xi}}}~\right)}}{\abs{\xi}^{1/4}\sqrt{\frac{\sqrt{\abs{\xi}}}{1-\abs{\xi}}+\arcsinh{\left(\sqrt{\frac{\abs{\xi}}{1-\abs{\xi}}}~\right)}}}~~,~\mathrm{when}~~\xi< 0 ~.
\end{equation}
%

%%%%%%%%%%%%%%%%%%%%%%%%%%%%%%%%%%%%%%%%%%%%%%%%%%%%%%%%%%%%%%%%%%%%%%%
\section{Imaginary part to leading-order in \texorpdfstring{$\xi$}{xi}}
\label{sec04}
%%%%%%%%%%%%%%%%%%%%%%%%%%%%%%%%%%%%%%%%%%%%%%%%%%%%%%%%%%%%%%%%%%%%%%%

The isotropic contribution is given by
\begin{equation} {\bf{Im}}~V_{(0)}({{r}}) =-g^2 C_F\int
\frac{d^3{\bf{p}}}{(2 \pi)^3} (e^{i{\bf{p\cdot r}}}-1)\frac{- \pi
 \lambda m_D^2}{p\,(p^2+m_D^2)^2} = - \frac{ g^2 C_F \lambda}{4 \pi } \,
\phi(\hat{r})~, \label{46}
\end{equation}
with
\begin{equation}
 \phi(\hat{r})= 2\int_0^{\infty}dz \frac{
z}{(z^2+1)^2} \left[1-\frac{\sin(z\, \hat{r})}{z\, \hat{r}}\right]~,
\label{47}
\end{equation}
and $\hat{r}\equiv r \,m_D$. This result has been derived before
in Refs.~\cite{Laine:2006ns,Laine:2007qy}. The term of order $\xi$
can be expressed as
\begin{eqnarray}
 {\bf{Im} }~\xi V_{(1)}({\bf{r}}) &=& -g^2 C_F \, \xi \int
\frac{d^3{\bf{p}}}{(2 \pi)^3} \, (e^{i{\bf{p \cdot r}}}-1) \,\nonumber\\
&\times&\left[\frac{3 \pi  \lambda m_D^2}{4
p\,(p^2+m_D^2)^2}\sin^2\alpha -\frac{2 \pi  \lambda m_D^4}{
p\,(p^2+m_D^2)^3}(\sin^2\alpha-\frac{1}{3})\right]\,\nonumber\\
&=&\frac{ g^2 C_F \xi \lambda}{4 \pi} \left[\psi_1(\hat{r},
\theta)+\psi_2(\hat{r}, \theta)\right]~,\label{48}
\end{eqnarray}
where $\theta$ is the angle between ${\bf{r}}$ and ${\bf{n}}$ and
\begin{eqnarray}
 \psi_1(\hat{r}, \theta) &=& \int_0^{\infty} dz
 \frac{z}{(z^2+1)^2}\left(1-\frac{3}{2}
 \left[\sin^2\theta\frac{\sin(z\, \hat{r})}{z\, \hat{r}}
 +(1-3\cos^2\theta)G(\hat{r}, z)\right]\right)\,~,\label{49}
\end{eqnarray}
\begin{eqnarray}
 \psi_2(\hat{r}, \theta) &=&- \int_0^{\infty} dz
\frac{\frac{4}{3}z}{(z^2+1)^3}\left(1-3 \left[
  \left(\frac{2}{3}-\cos^2\theta \right) \frac
 {\sin(z\, \hat{r})}{z\, \hat{r}}+(1-3\cos^2\theta)
 G(\hat{r},z)\right]\right)~,\label{410}
\end{eqnarray}
with
\begin{equation}
 G(\hat{r}, z)= \frac{\hat{r} z\cos(\hat{r} z)- \sin(\hat{r} z)
 }{(\hat{r} z)^3}~.
\label{411}
\end{equation}

Note that
\begin{eqnarray}
 \phi(\hat{r})&=& -\frac{1}{9}\, \hat{r}^2(-4+3\gamma_E+3\ln
 \hat{r})~,\nonumber \\
\psi_1(\hat{r}, \theta)&=&\frac{1}{600}\,
\hat{r}^2[123-90\gamma_E-90\ln
\hat{r}+\cos(2\theta)(-31+30\gamma_E+30\ln \hat{r})]~,\nonumber
\\\psi_2(\hat{r},
\theta)&=&\frac{1}{90}\,\hat{r}^2(-4+3\cos(2\theta))~,\label{51}
\end{eqnarray}
where $\gamma_E$ is the Euler-Gamma constant.

%%%%%%%%%%%%%%%%%%%%%%%%%%%%%%%%%%%%%%%%%%%%%%%%%%%%%%%%%%%%%%
\subsection{Angle-averaged imaginary part to leading-order in small~-~\texorpdfstring{$\xi$}{xi}~limit} 
\la{sect:loxiangleavg}
%%%%%%%%%%%%%%%%%%%%%%%%%%%%%%%%%%%%%%%%%%%%%%%%%%%%%%%%%%%%%%

From \cite{Dumitru:2009fy}, we can write down the imaginary part of the potential 
\begin{eqnarray}
V_I(r)&=& Im V_{(0)}(r)+Im \,\xi V_{(1)}(r)\nonumber\\
&=& -\alpha_s C_F \lambda \phi (\hat{r})+\alpha_s C_F \lambda \xi \left[\psi_1(\hat{r},\theta)+\psi_2(\hat{r},\theta)\right]~,
\end{eqnarray}

where, $\hat{r}\equiv r \,m_D$, $\theta$ is the angle between ${\bf{r}}$ and ${\bf{n}}$ (direction of the anisotropy \cite{Romatschke:2003ms}), and
\begin{equation}
\phi(\hat{r})= 2\int_0^{\infty}dz \frac{
	z}{(z^2+1)^2} \left[1-\frac{\sin(z\, \hat{r})}{z\, \hat{r}}\right]~,\label{1.9}
\end{equation}
\begin{equation}
\psi_1(\hat{r}, \theta) = \int_0^{\infty} dz
\frac{z}{(z^2+1)^2}\left[1-\frac{3}{2}
\left\{\sin^2\theta\frac{\sin(z\, \hat{r})}{z\, \hat{r}}
+(1-3\cos^2\theta)G(\hat{r}, z)\right\}\right]~,
\end{equation}
\begin{equation}
\psi_2(\hat{r}, \theta) =- \int_0^{\infty} dz
\frac{\frac{4}{3}z}{(z^2+1)^3}\left[1-3 \left\{
\left(\frac{2}{3}-\cos^2\theta \right) \frac
{\sin(z\, \hat{r})}{z\, \hat{r}}+(1-3\cos^2\theta)
G(\hat{r},z)\right\}\right]~,
\end{equation}
with
\begin{equation}
G(\hat{r}, z)= \frac{\hat{r} z\cos(\hat{r} z)- \sin(\hat{r} z)
}{(\hat{r} z)^3}~.
\end{equation}

Now, angle-averaged $\psi$-functions are obtained as 
\begin{eqnarray}
\langle \psi_1\rangle_{\theta}&=&\frac{\int_{0}^{\pi}\psi_1(\hat{r}, \theta)\sin\theta d\theta}{\int_{0}^{\pi}\sin\theta d\theta}\nonumber\\
&=&\int_0^{\infty} dz
\frac{z}{(z^2+1)^2}\left[1-\frac{\sin(z\, \hat{r})}{z\, \hat{r}}\right]\nonumber\\
&=&\frac{1}{2}\phi(\hat{r})\label{1.13}
\end{eqnarray}
and
\begin{eqnarray}
\langle \psi_2\rangle_{\theta}&=&\frac{\int_{0}^{\pi}\psi_2(\hat{r}, \theta)\sin\theta d\theta}{\int_{0}^{\pi}\sin\theta d\theta}\nonumber\\
&=&-\int_0^{\infty} dz
\frac{\frac{4}{3}z}{(z^2+1)^3}\left[1-\frac{\sin(z\, \hat{r})}{z\, \hat{r}}\right]\nonumber\\
&=&-\frac{4}{3}\int_0^{\infty} dz
\frac{z}{(z^2+1)^3}\left[1-\frac{\sin(z\, \hat{r})}{z\, \hat{r}}\right]\label{1.14}
\end{eqnarray}
Eq.(\ref{1.9}) can be simplified as 
\begin{eqnarray}
\phi(\hat{r})&=& 2\int_0^{\infty}dz \frac{z}{(z^2+1)^2} \left[1-\frac{\sin(z\, \hat{r})}{z\, \hat{r}}\right]\nonumber\\
&=&2\int_0^{\infty}dz \frac{z}{(z^2+1)^2}-2\int_0^{\infty}dz \frac{z}{(z^2+1)^2}.\frac{\sin(z\, \hat{r})}{z\, \hat{r}}\nonumber\\
&=&2.\frac{1}{2}-2\int_0^{\infty}dz \frac{z}{(z^2+1)^2}.\frac{\sin(z\, \hat{r})}{z\, \hat{r}}\label{1.15}\\
&=&1-2\int_0^{\infty}dz \frac{z}{(z^2+1)^2}.\frac{\sin(z\, \hat{r})}{z\, \hat{r}}\label{1.16}\\
&=&1-2\int_{0}^{\infty}\frac{dz}{\hat{r}}.\frac{\frac{z}{\hat{r}}}{\left(\frac{z^2}{\hat{r}^2}+1\right)^2}.\frac{\sin(z)}{z}\label{1.17}\\
&=& 1-2\hat{r}^2\int_{0}^{\infty}dz~\frac{1}{\left(z^2+\hat{r}^2\right)^2}~\sin(z)~,
\end{eqnarray}
where in Eq.(\ref{1.15}) we use $\int_0^{\infty}dz \frac{z}{(z^2+1)^2}=\frac{1}{2}$, and in going from Eq.(\ref{1.16}) to Eq.(\ref{1.17}) we use change of variables : $\left\{z\hat{r},z,dz\right\}=\left\{z,\frac{z}{\hat{r}},\frac{dz}{\hat{r}}\right\}$.

Similarly, Eq.(\ref{1.14}) can be simplified as 
\begin{eqnarray}
\langle \psi_2\rangle_{\theta}&=&-\frac{4}{3}\int_0^{\infty} dz
\frac{z}{(z^2+1)^3}\left[1-\frac{\sin(z\, \hat{r})}{z\, \hat{r}}\right]\nonumber\\
&=&-\frac{4}{3}\left[\int_0^{\infty} dz\frac{z}{(z^2+1)^3}-\int_0^{\infty} dz\frac{z}{(z^2+1)^3}.\frac{\sin(z\, \hat{r})}{z\, \hat{r}}\right]\nonumber\\
&=&-\frac{4}{3}\left[\frac{1}{4}-\int_0^{\infty} dz\frac{z}{(z^2+1)^3}.\frac{\sin(z\, \hat{r})}{z\, \hat{r}}\right]\label{1.19}\\
&=&-\frac{4}{3}\left[\frac{1}{4}-\int_{0}^{\infty}\frac{dz}{\hat{r}}.\frac{\frac{z}{\hat{r}}}{\left(\frac{z^2}{\hat{r}^2}+1\right)^3}.\frac{\sin(z)}{z}\right]\label{1.20}\\
&=&-\frac{4}{3}\left[\frac{1}{4}-\hat{r}^4\int_{0}^{\infty}dz~\frac{1}{\left(z^2+\hat{r}^2\right)^3}~\sin(z)\right]~,
\end{eqnarray}
where in Eq.(\ref{1.19}) we use $\int_0^{\infty}dz \frac{z}{(z^2+1)^3}=\frac{1}{4}$, and in going from Eq.(\ref{1.19}) to Eq.(\ref{1.20}) we use change of variables : $\left\{z\hat{r},z,dz\right\}=\left\{z,\frac{z}{\hat{r}},\frac{dz}{\hat{r}}\right\}$.

Hence, in summary, angle-averaged imaginary part of the potential to leading-order in $\xi$ is given by 
\begin{equation}
\langle V_I(r)\rangle_{\theta}=-\alpha_s C_F \lambda \phi (\hat{r})+\alpha_s C_F \lambda \xi \left[\langle \psi_1\rangle_{\theta}+\langle \psi_2\rangle_{\theta}\right]~,\label{1.22}
\end{equation}
where,
\begin{equation}
\phi(\hat{r})=1-2\hat{r}^2\int_{0}^{\infty}dz~\frac{1}{\left(z^2+\hat{r}^2\right)^2}~\sin(z)~,
\end{equation}
\begin{equation}
\langle \psi_1\rangle_{\theta}=\frac{1}{2}\phi(\hat{r})~,\label{1.24}
\end{equation}
and
\begin{equation}
\langle\psi_2\rangle_{\theta}=-\frac{4}{3}\left[\frac{1}{4}-\hat{r}^4\int_{0}^{\infty}dz~\frac{1}{\left(z^2+\hat{r}^2\right)^3}~\sin(z)\right]~.
\end{equation}
Now we will find an expression for $\langle\psi_2\rangle_{\theta}$ in terms of $\phi(\hat{r})$ and it's first derivative i.e., $\phi'(\hat{r})$.
\begin{eqnarray}
\phi(\hat{r})&=&1-2\hat{r}^2\int_{0}^{\infty}dz~\frac{1}{\left(z^2+\hat{r}^2\right)^2}~\sin(z)\nonumber\\
\Rightarrow \phi'(\hat{r}) &=& 0-4\hat{r}\int_{0}^{\infty}dz~\frac{1}{\left(z^2+\hat{r}^2\right)^2}~\sin(z)+2\hat{r}^2\int_{0}^{\infty}dz~\frac{4\hat{r}}{\left(z^2+\hat{r}^2\right)^3}~\sin(z)\nonumber\\
\Rightarrow \phi'(\hat{r}) &=&-4\hat{r}\int_{0}^{\infty}dz~\frac{1}{\left(z^2+\hat{r}^2\right)^2}~\sin(z)+8\hat{r}^3\int_{0}^{\infty}dz~\frac{1}{\left(z^2+\hat{r}^2\right)^3}~\sin(z)\nonumber\\
\Rightarrow \frac{1}{6}\hat{r}\phi'(\hat{r}) &=&-\frac{2}{3}\hat{r}^2\int_{0}^{\infty}dz~\frac{1}{\left(z^2+\hat{r}^2\right)^2}~\sin(z)+\frac{4}{3}\hat{r}^4\int_{0}^{\infty}dz~\frac{1}{\left(z^2+\hat{r}^2\right)^3}~\sin(z)\nonumber
\end{eqnarray}
\begin{multline*}
\Rightarrow -\frac{1}{3}\phi(\hat{r})+\frac{1}{6}\hat{r}\phi'(\hat{r}) =-\frac{1}{3}+\frac{2}{3}\hat{r}^2\int_{0}^{\infty}dz~\frac{1}{\left(z^2+\hat{r}^2\right)^2}~\sin(z)\\-\frac{2}{3}\hat{r}^2\int_{0}^{\infty}dz~\frac{1}{\left(z^2+\hat{r}^2\right)^2}~\sin(z)+\frac{4}{3}\hat{r}^4\int_{0}^{\infty}dz~\frac{1}{\left(z^2+\hat{r}^2\right)^3}~\sin(z)
\end{multline*}
\begin{eqnarray}
\Rightarrow -\frac{1}{3}\phi(\hat{r})+\frac{1}{6}\hat{r}\phi'(\hat{r}) &=&-\frac{1}{3}+\frac{4}{3}\hat{r}^4\int_{0}^{\infty}dz~\frac{1}{\left(z^2+\hat{r}^2\right)^3}~\sin(z)\nonumber\\
&=&-\frac{4}{3}\left[\frac{1}{4}-\hat{r}^4\int_{0}^{\infty}dz~\frac{1}{\left(z^2+\hat{r}^2\right)^3}~\sin(z)\right]\nonumber\\
&=&\langle\psi_2\rangle_{\theta}
\end{eqnarray}
Hence, we get 
\begin{eqnarray}
\langle\psi_2\rangle_{\theta} &=& -\frac{1}{3}\phi(\hat{r})+\frac{1}{6}\hat{r}\phi'(\hat{r})\nonumber\\
&=&-\frac{1}{3}\left[\phi(\hat{r})-\frac{1}{2}\hat{r}\phi'(\hat{r})\right]~.\label{1.27}
\end{eqnarray}
Inserting Eq.(\ref{1.24}) and Eq.(\ref{1.27}) in Eq.(\ref{1.22}), we obtain angle-averaged imaginary part of the potential
\begin{eqnarray}
\langle V_I(r)\rangle_{\theta}&=& -\alpha_s C_F \lambda \phi (\hat{r})+\alpha_s C_F \lambda \xi \left[\big < \psi_1\big>_{\theta}+\big < \psi_2\big>_{\theta}\right]\nonumber\\
&=&-\alpha_s C_F \lambda \phi (\hat{r})+\alpha_s C_F\lambda\xi
\left[\frac{1}{2}\phi(\hat{r})-\frac{1}{3}\phi(\hat{r})+\frac{1}{6}\hat{r}\phi'(\hat{r})\right]\nonumber\\
&=&-\alpha_s C_F \lambda \phi (\hat{r})+\alpha_s C_F\lambda\xi
\left[\frac{1}{6}\phi(\hat{r})+\frac{1}{6}\hat{r}\phi'(\hat{r})\right]\nonumber\\
&=&-\alpha_s C_F \lambda \phi (\hat{r})+\alpha_s C_F\lambda\xi
\frac{1}{6}\phi(\hat{r})+\alpha_s C_FT\xi\frac{1}{6}\hat{r}\phi'(\hat{r})\nonumber\\
&=&-\alpha_s C_F \lambda \phi (\hat{r})\left[1-\frac{\xi}{6}\right]+\alpha_s C_F\lambda\xi\frac{1}{6}\hat{r}\phi'(\hat{r})\nonumber\\
&=&-\alpha_s C_F \lambda\left[\left(1-\frac{\xi}{6}\right)\phi (\hat{r})-\frac{\xi}{6}\hat{r}\phi'(\hat{r})\right]
\end{eqnarray}

In our model we assume $\langle\psi_2\rangle_{\theta} \simeq -\frac{1}{3}\phi(\hat{r})$, and Eq.(\ref{1.22}) becomes 
\begin{eqnarray}
\langle V_I(r)\rangle_{\theta}&=& -\alpha_s C_F \lambda \phi (\hat{r})+\alpha_s C_F \lambda \xi \left[\big < \psi_1\big>_{\theta}+\big < \psi_2\big>_{\theta}\right]\nonumber\\
&\simeq&-\alpha_s C_F \lambda \phi (\hat{r})+\alpha_s C_F\lambda\xi
\left[\frac{1}{2}\phi(\hat{r})-\frac{1}{3}\phi(\hat{r})\right]\nonumber\\
&\simeq&-\alpha_s C_F \lambda \phi (\hat{r})+\alpha_s C_F\lambda\xi~\frac{1}{6}\phi(\hat{r})\nonumber\\
&\simeq&-\alpha_s C_F \lambda \phi (\hat{r})\left[1-\frac{\xi}{6}\right] .
\end{eqnarray}
We take the final line above as our approximate model of the small $\xi$ behavior
\begin{equation}
\lim_{\xi \rightarrow 0} \langle V_I(r)\rangle^{\mathrm{model}}_{\theta}= -\alpha_s C_F \lambda \phi (\hat{r}) \left[ 1 - \frac{\xi}{6} \right] .
\label{eq:weakanisoimv}
\end{equation}

\subsection{Imaginary part to leading-order in large~-~\texorpdfstring{$\xi$}{xi}~limit}

We have the static Feynman propagator \cite{Nopoush:2017zbu}
\ba
&&\lim_{\omega\rightarrow 0}\tilde{\cal D}^{00}_{F}(p,\xi)= \nonumber\\
&&\hspace{7mm} \frac{4i \lambda m_D^2}{\varsigma{|\bf p}|[({\bf p}^2+m_\beta^2)({\bf p}^2 + m_\alpha^2 + m_\gamma^2)-m_\delta^4]^2}
\left[ \frac{m_\delta^4- \varsigma ( {\bf p}^2+m_\alpha^2 + m_\gamma^2)^2 }{1+\varsigma}  E\!\left(-\varsigma \right) -
m_\delta^4 K\!\left(-\varsigma \right) 
\right]\!, \hspace{12mm} \label{eq:term2alt}
\ea
where $\varsigma \equiv \xi p_x^2/{\bf p}^2$, $\lambda$ is a temperature-like scale which corresponds to temperature in the equilibrium limit, and $K$ and $E$ are complete elliptic integrals of the first and second kind, respectively, defined by 
\ba
K(x)&\equiv& \int _0^{\pi/2}\frac{1}{\sqrt{1-x\sin^2\phi}} \,d\phi\,,\nonumber \\
E(x)&\equiv& \int _0^{\pi/2}\sqrt{1-x\sin^2\phi}\,\,d\phi\,.
\ea
And the masses $m^2_{\alpha,\beta,\gamma,\delta}$ are given by in the static limit \cite{Romatschke:2003ms,Dumitru:2007hy}
\ba
m_\alpha^2&\equiv&-\frac{m_D^2}{2 p_x^2 \sqrt{\xi}}%
\left(p_z^2 {\rm{arctan}}{\sqrt{\xi}}-\frac{p_z {\bf{p}}^2}{\sqrt{{\bf{p}}^2+\xi p_x^2}}%
{\rm{arctan}}\bigg[\frac{\sqrt{\xi} p_{z}}{\sqrt{{\bf{p}}^2+\xi p_x^2}}\bigg]\right) \; , \label{eq:malpha}\\
m_\beta^2&\equiv&m_{D}^2
\frac{(\sqrt{\xi}+(1+\xi){\rm{arctan}}{\sqrt{\xi}})({\bf{p}}^2+\xi p_x^2)+\xi p_z\left(%
	p_z \sqrt{\xi} + \frac{{\bf{p}}^2(1+\xi)}{\sqrt{{\bf{p}}^2+\xi p_x^2}} %
	{\rm{arctan}}\Big[\frac{\sqrt{\xi} p_{z}}{\sqrt{{\bf{p}}^2+\xi p_x^2}}\Big]\right)}{%
	2  \sqrt{\xi} (1+\xi) ({\bf{p}}^2+ \xi p_x^2)}\,, \;\;\;\; \hspace{12mm}\label{eq:mbeta} \\
m_\gamma^2&\equiv&-\frac{m_D^2}{2}\left(\frac{{\bf{p}}^2}{{\bf{p}}^2 + \xi p_x ^2}%
-\frac{{\bf p}^2+p_z^2}{\sqrt{\xi}p_x^2}{\rm{arctan}}{\sqrt{\xi}}+\frac{
	p_z{\bf{p}}^2(2{\bf{p}}^2+3\xi p_x^2)}{\sqrt{\xi}(\xi
	p_x^2+{\bf{p}}^2)^{{3/2}}
	p_x^2}{\rm{arctan}}\bigg[\frac{\sqrt{\xi}
	p_{z}}{\sqrt{{\bf{p}}^2+\xi p_x^2}}\bigg]\right) , \label{eq:mgamma} \\
m_\delta^4&\equiv&\frac{\pi^2 m_D^4 \xi^2 p_z^2 p_x^2 {\bf{p}}^2}{16({\bf{p}}^2 + \xi p_x ^2)^{3}}\, . \label{eq:mdelta}
\ea
\checked{mn}
Here, ${\bf n}=(0,0,1)$ points along the $z$-axis and ${\bf p}$ lies in the $x-z$ plane; in the general case, one should take  $p_z\rightarrow \bf{p\cdot n}$ and $p_x\rightarrow |\bf{p- (p\cdot n)n}|$.\\
With this, we can write down an expression for the real part of the potential which is valid to all orders in $\xi$ \cite{Dumitru:2007hy,Strickland:2011aa}
\ba
V({\bf{r}},\xi) &=& -g^2 C_F\int \frac{d^3{\bf{p}}}{(2\pi)^3} \,
\left( e^{i{\bf{p \cdot r}}} -1 \right) \tilde{\cal D}_R^{00}(\omega=0, \bf{p},\xi) \nonumber\\
&=& -g^2 C_F\int \frac{d^3{\bf{p}}}{(2\pi)^3} \,
\left( e^{i{\bf{p \cdot r}}} -1 \right) \frac{{\bf{p}}^2+m_\alpha^2+m_\gamma^2}
{({\bf{p}}^2 + m_\alpha^2 +
	m_\gamma^2)({\bf{p}}^2+m_\beta^2)-m_\delta^4} \, ,
\label{eq:repot}
\ea

\paragraph*{}
Expanding Eq.(\ref{eq:term2alt}) in terms of powers of $\xi$ in the large-$\xi$ limit, the static Feynman propagator takes the form

\begin{equation}
\lim_{\omega\rightarrow 0} \tilde{\cal D}^{00}_{F}(p,\xi) = -\frac{4im^2_D\lambda\sqrt{\frac{1}{\xi}}}{p^4\,p_x}+\frac{2im^4_D\pi\lambda}{p^6\,p_x\,\xi}+{\cal O}\left(\frac{1}{\xi}\right)^{3/2} \, ,\label{2.38}
\end{equation}
The imaginary part of the heavy-quark potential can be obtained from the Fourier transform of the static limit of the Feynman propagator \cite{Nopoush:2017zbu}
\be
V_I({\bf r},\xi) \equiv - \frac{g^2 C_F}{2} \int \frac{d^3{\bf p}}{(2\pi)^3} \left(e^{i {\bf p}\cdot {\bf r}} - 1 \right) 
\tilde{\cal D}^{00}_F \Big|_{\omega \rightarrow 0} \; .
\ee
From this we learn that in the limit $\xi \rightarrow \infty$ the magnitude of the imaginary part of the potential will decrease as $1/\sqrt{\xi}$.

\subsection{Model for the imaginary part of the potential for general~\texorpdfstring{$\xi$}{xi}}

In order to construct a model that reduces to our model potential in the weak anisotropy limit \eqref{eq:weakanisoimv} and is proportional to $1/\sqrt{\xi}$ in the large-$\xi$ limit, we construct the following ansatz
\begin{equation}
\langle V_I(r) \rangle^{\mathrm{model}}_{\theta} = - \frac{\alpha_s C_F \lambda \phi (\hat{r})}{ \sqrt{1 + \frac{\xi}{3}} } \, ,
\label{eq:allanisoimv}
\end{equation}
with $\hat{r} = m_D(\lambda) r$ and $m_D(\lambda)=F_{m_D}\lambda~\sqrt{N_c\left(1+\frac{N_F}{6}\right)\pi\alpha}$ .

One can also verify that in the small-$\xi$ limit i.e., 
\begin{equation}
\lim_{\xi\rightarrow 0}\frac{1}{\sqrt{1+\frac{\xi}{3}}}=1-\frac{\xi}{6}+{\cal O}\left(\xi\right)^2\, ,
\end{equation}
our model potential \eqref{eq:allanisoimv} yields  potential in the weak anisotropy limit \eqref{eq:weakanisoimv} and in the large-$\xi$ limit i.e.,  
\begin{equation}
\lim_{\xi\rightarrow \infty}\frac{1}{\sqrt{1+\frac{\xi}{3}}}=\sqrt{3}~\sqrt{\frac{1}{\xi}}+{\cal O}\left(\frac{1}{\xi}\right)^{3/2}\, ,
\end{equation}
our model potential \eqref{eq:allanisoimv} is proportional to $1/\sqrt{\xi}$ which justifies Eq.(\ref{2.38}).

%%%%%%%%%%%%%%%%%%%%%%%%%%%%%%%%%%%%%%%%%%%%%%%%%%%%%%%%%%%%%%
\section{The complete anisotropic complex potential}
\label{sec05}
%%%%%%%%%%%%%%%%%%%%%%%%%%%%%%%%%%%%%%%%%%%%%%%%%%%%%%%%%%%%%%
Similarly, as we did for the isotropic case, to match smoothly onto the zero temperature limit, we use
\begin{equation}
V^{\rm aniso}_R (r)=
\begin{cases}  V^{\rm aniso}_{\rm KMS}(r)  &\mbox{if } V^{\rm aniso}_{\rm KMS}(r)  \leq V_{\rm vac}(r_{\rm SB}) \\
V_{\rm vac}(r_{\rm SB}) & \mbox{if } V^{\rm aniso}_{\rm KMS}(r) > V_{\rm vac}(r_{\rm SB})
\end{cases} \, .
\label{eq:vmedre}
\end{equation}
In the limit $T\rightarrow0$, Eq.~\eqref{eq:vmedre} reduces to Eq.~\eqref{eq:vvac}.

For the imaginary part of the potential, we use Eq. \eqref{eq:allanisoimv}
\begin{equation}
V^{\rm aniso}_I (r)= \langle V_I(r) \rangle^{\mathrm{model}}_{\theta}=- \frac{\alpha_s C_F \lambda \phi (\hat{r})}{ \sqrt{1 + \frac{\xi}{3}} }\,.
\label{eq:vmedim}
\end{equation}

The resulting final anisotropic complex-valued potential is of the form 
\begin{equation}
V^{\rm aniso} (r) = V^{\rm aniso}_R (r) + i V^{\rm aniso}_I (r) \, .
\label{eq:vform}
\end{equation}
As discussed in Sec. \ref{intro}, the phenomenological anisotropic complex-valued potential given in Eq. (\ref{eq:vform}), together with Eqs.~\eqref{eq:vvac}~-~\eqref{eq:vmedim}, is referred to as the anisotropic (Aniso) model. After solving the resulting evolution using the quantum-trajectories (QTraj) framework, we refer to the corresponding predictions as ``QTraj-Aniso'' throughout this work.
%%%%%%%%%%%%%%%%%%%%%%%%%%%%%%%%%%%%%%%%%%%%%%%%%%%%%%%%%%%%%%%%%%%%%%%%%
\section{Numerical method for solving the Schr\"odinger equation}
\label{sec06}
%%%%%%%%%%%%%%%%%%%%%%%%%%%%%%%%%%%%%%%%%%%%%%%%%%%%%%%%%%%%%%%%%%%%%%%
Since the complex potential is spherically symmetric, the general solution in spherical coordinates can be expressed as
\be
\psi(r,\theta,\phi,t) = \sum_{\ell m} R_{\ell m}(r,t) Y_{\ell m}(\theta,\phi) \, ,
\ee
where $Y_{\ell m}$ denote the standard spherical harmonics. A subsequent change of variable, defined by \mbox{$u_{\ell m}(r,t) \equiv r R_{\ell m}(r,t)$}, simplifies the radial equation. This leads to the expression
\be
u(r,\theta,\phi,t) = \sum_{\ell m} u_{\ell m}(r,t) Y_{\ell m}(\theta,\phi) \, ,
\ee
with $u(r,\theta,\phi,t) = r \psi(r,\theta,\phi,t)$. The transformation results in an effective one-dimensional Hamiltonian for each angular momentum channel $\ell$:
\be
\hat{H}_\ell = \frac{\hat{p}^2}{2m} + V_{{\rm eff},\ell}(r,t) \, ,
\ee
where the effective potential is $V_{{\rm eff},\ell}(r,t) = V(r,t) + \frac{\ell(\ell+1)}{2 m r^2}$  and the radial momentum operator is $\hat{p} = -i \partial/\partial r$.

\subsection{Time Evolution}

As shown in \cite{Boyd:2019arx}, applying the time-evolution operator to $u$ yields
\ba
u(r,\theta,\phi,t + \Delta t) &=& \exp(- i \hat{H} \Delta t) u(r,\theta,\phi,t ) \nonumber \\
&=&  {\cal N} \sum_{\ell,m} \frac{1}{\sqrt{2\ell+1}} Y_{\ell m}(\theta,\phi) \exp(- i \hat{H}_\ell \Delta t) u_\ell(r,t)  \, ,
\ea
where ${\cal N}$ is a normalization constant and summation limits are implicit. Utilizing the projection $u_{\ell}(r,t) = \sqrt{2\ell+1} \int d\Omega \, Y_{\ell m}^*(\theta,\phi) \, u(r,\theta,\phi,t)$, the update rule for each radial function $u_\ell$ is given by
\be
u_\ell(r,t + \Delta t) = \exp(- i \hat{H}_\ell \Delta t) u_\ell(r,t) \, .
\label{eq:uUpdate}
\ee
This demonstrates that the time evolution for each partial wave $\ell$ proceeds independently under its own Hamiltonian $\hat{H}_\ell$. The wave functions for each $\ell$ are normalized such that
\be
\int_0^\infty dr \, |u_\ell(r,t)|^2 = 1 \, .
\ee

\subsection{Numerical Implementation}

We will use the time evolution operator \eqref{eq:uUpdate} to evolve the wave function with a
particular initial condition. The boundary condition $u_\ell(0,t) = 0$ is automatically enforced by expanding the real and imaginary parts of the wave function in a basis of Fourier sine series. The evolution over a single timestep $\Delta t$ is implemented through the following steps \cite{Boyd:2019arx}:

%%%

\begin{itemize}
	\item[\bfseries 1.] Configuration space half-step: Apply the phase rotation from the effective potential: $ \psi_1 = \exp(- i V_{{\rm eff},\ell} \Delta t/2) \psi_0$.
	
	\item[\bfseries 2.] Momentum space transformation: Compute the Fourier sine transform ($\mathbb{F}_s$) of the real and imaginary components: $\tilde\psi_1 = \mathbb{F}_s[\Re \psi_1] + i \mathbb{F}_s[\Im \psi_1] $.
	
	\item[\bfseries 3.] Momentum space update: Apply the phase rotation from the kinetic energy operator: $\tilde\psi_2 =  \exp\!\left(-i \frac{p^2}{2 m} \Delta t\right) \tilde\psi_1$.
	
	\item[\bfseries 4.] Configuration space transformation: Compute the inverse Fourier sine transform ($\mathbb{F}_s^{-1}$): $\psi_2 = \mathbb{F}_s^{-1}[\Re \tilde\psi_2] + i \mathbb{F}_s^{-1}[\Im \tilde\psi_2] $.
	
	\item[\bfseries 5.] Configuration space half-step: Apply the final phase rotation from the effective potential: $ \psi_3 = \exp(- i V_{{\rm eff},\ell} \Delta t/2) \psi_2$.
\end{itemize}

%%%
The Discrete Sine Transforms (DST) above are efficiently computed using Fast Fourier Transform (FFT) routines \cite{Boyd:2019arx}. For our simulations, the algorithm is massively parallelized on GPUs using the optimized CUDA CUFFT library \cite{cuda}.

%%%%%%%%%%%%%%%%%%%%%%%%%%%%%%%%%%%%%%%%%%%%%%%%%%%%%%%%%%%%%%
\section{Computation of \texorpdfstring{$R_{AA}$}{RAA} and \texorpdfstring{$v_{2}$}{v2}~including feed-down}
\label{sec07}
%%%%%%%%%%%%%%%%%%%%%%%%%%%%%%%%%%%%%%%%%%%%%%%%%%%%%%%%%%%%%

Using the complex potential of Eq.~\eqref{eq:vform} together with Eqs.~\eqref{eq:vmedre} and \eqref{eq:vmedim}, we solve the time-dependent Schr\"odinger equation numerically on a discrete radial lattice. The evolution is carried out with a split-step pseudospectral scheme based on discrete sine transforms (DST), which is unitary for real potentials \cite{Fornberg:1978,TAHA1984203,Boyd:2019arx}. We discretize the system with $N=4096$ points up to $r_{\rm max}=19.7$ fm, corresponding to a spacing $a\simeq 0.0048$ fm. The in-medium suppression is computed for $\ell=0$ and $\ell=1$ states separately using Eq.~\eqref{eq:uUpdate}. As initial condition we employ a Gaussian profile \cite{Islam:2020gdv}
\be
u_\ell(r,\tau=0) \propto r^{\ell+1} \exp(-r^2/\Delta^2) \, ,
\ee
with $\Delta=0.04$ fm, which effectively mimics local bottomonium production from hard scatterings. The survival probability of a given vacuum state is then obtained from the overlap of the evolved in-medium wave function with the corresponding vacuum basis state.

We solve the 3+1D Schrödinger equation for quarkonium states on a realistic 3+1D hydrodynamics background (aHydroQP) tuned to 5.02 TeV data \cite{Alqahtani:2020paa,Alqahtani:2017mhy}. The background uses smooth optical Glauber initial conditions with an initial central temperature $T_0 = 630$ MeV at $\tau_0 = 0.25$ fm/c and a specific shear viscosity of $4\pi\eta/s = 2$~\cite{Alqahtani:2020paa}. Wave-packets are evolved with the vacuum potential until a medium formation time $\tau_\text{med}$, after which the in-medium complex potential is applied. The vacuum potential is reinstated when the local temperature drops below $T_\text{QGP} = 155$ MeV.

We numerically solve the time-dependent Schrödinger equation for a large set of bottomonium trajectories (1 million) to account for varying local temperatures. Initial production points are sampled from the nuclear binary overlap profile, $N_{AA}^\text{bin}(x,y)$. All states are generated at mid-rapidity ($y=0$) with transverse momenta sampled from a $p_T/(p_T^2 + \langle M \rangle^2)^2$ distribution and random azimuthal angles $\phi \in [0, 2\pi]$. For each sampled state, the QGP temperature along its straight-line trajectory is extracted from the aHydroQP background.

After propagating each state along its trajectory, the survival probabilities are converted into particle yields by multiplying with (i) the average number of binary collisions in the selected centrality bin and (ii) the primordial production cross section of each state.  

To incorporate late-time feed down, we introduce a feed-down matrix $F$ built from Particle Data Group branching fractions~\cite{ParticleDataGroup:2020ssz}:
\begin{equation}
	F_{ij} = \left\{ \begin{matrix}
		\text{branching fraction $j$ to $i$}, & \text{for } i < j, \\
		1, & \text{for } i = j, \\
		0, & \text{for } i > j,
		\end{matrix} \right.\label{eq:fdmnew}
\end{equation}
which accounts for decays of excited states into lower states (see also eq.~(6.4) of ref.~\cite{Brambilla:2020qwo}). The calculation of different elements of the feed-down matrix \eqref{eq:fdmnew} is demonstrated in \cite{Islam:2020bnp}.

In the case of p-p collisions one can take the primordial cross sections for production and convert this into the post feed down cross sections using $\vec{\sigma}_\text{exp} = F \vec{\sigma}_\text{primordial}$. Our considered states are $\vec{\sigma} = \{ \Upsilon(1S),\,$ $\Upsilon(2S),\,$ $\chi_{b0}(1P),\,$ $\chi_{b1}(1P),\,$ $\chi_{b2}(1P),\,$ $\Upsilon(3S),\,$ $\chi_{b0}(2P),\,$ $\chi_{b1}(2P),\,$ $\chi_{b2}(2P)\}$. Once we know the experimental values for the production cross-sections $\vec{\sigma}_{\text{exp}}=\{57.6$, 19, 3.72, 13.69, 16.1, 6.8, 3.27, 12.0, $14.15\}$ nb, one can compute the primordial cross sections $\vec{\sigma}_{\text{primordial}}$
%=\{37.97$, 18.27, 3.79, 20.88, 19.53, 8.21, 3.33, 17.28, $17.07\}$ nb 
via $\vec{\sigma}_\text{primordial} = F^{-1} \vec{\sigma}_\text{exp}$.
The final observable yields in heavy-ion collisions are obtained by applying the feed-down matrix to the simulated post-QGP particle numbers: $\vec{N}_\text{final} = F \vec{N}_\text{QGP}$, where $\vec{N}_\text{QGP}$ is constructed for each state as (survival probability) $\times \langle N_\text{bin}(b) \rangle \times \sigma_\text{primordial}$.

We compute $R_{AA}$ for each state by dividing the final post–feed-down yield by the product of the average number of binary collisions in the sampled centrality class and the corresponding post–feed-down $pp$ cross section ($\sigma_\text{exp}$). The momentum anisotropy of bottomonia is obtained analogously. Since the aHydroQP background has the reaction plane aligned with the $x$–$y$ axes ($\Psi_\text{RP}=0$), we evaluate
$v_n = \langle \cos(n\phi) \rangle $, averaging over all post–feed-down particles in a given centrality and $p_T$ bin. For both $R_{AA}$ and $v_2$, statistical uncertainties are estimated directly from the ensemble of sampled quantum trajectories.

%%%%%%%%%%%%%%%%%%%%%%%%%%%%%%%%%%%%%%%%%%%%%%%%%%%%%%%%%%%%%%
\section{Results}
\label{sec08}
%%%%%%%%%%%%%%%%%%%%%%%%%%%%%%%%%%%%%%%%%%%%%%%%%%%%%%%%%%%%%%
In this section, we will present our results for the nuclear modification factor ($R_{AA}$), double ratios, and the elliptic flow ($v_2$) of $\Upsilon(1S)$, $\Upsilon(2S)$, and $\Upsilon(3S)$ states.
%%%%%%%%%%%%%%%%%%%%%%%%%%%%%%%%%%%%%%%%%%%%%%%%%%%%%%%%%%%%%%
\subsection{Nuclear suppression factor~\texorpdfstring{$R_{AA}$}{RAA}~and double ratios}
\label{sec:raa-doubleratio}
%%%%%%%%%%%%%%%%%%%%%%%%%%%%%%%%%%%%%%%%%%%%%%%%%%%%%%%%%%%%%%

%--------------------------
\begin{figure}[ht]
	\begin{center}
		\includegraphics[width=0.485\linewidth]{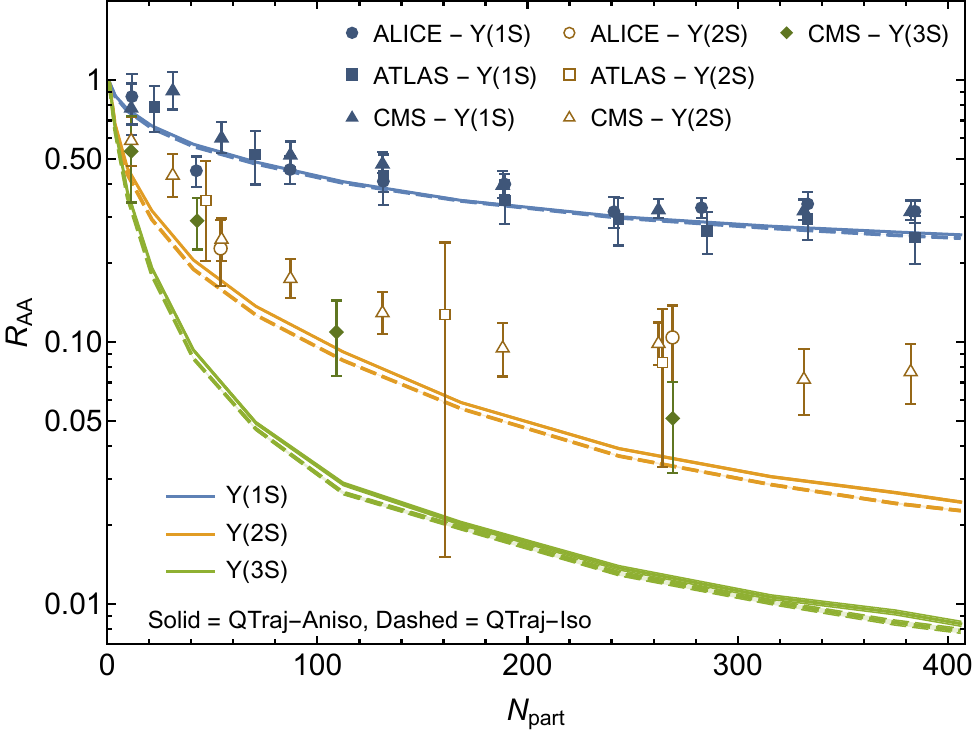}\hspace{2mm}
		\includegraphics[width=0.485\linewidth]{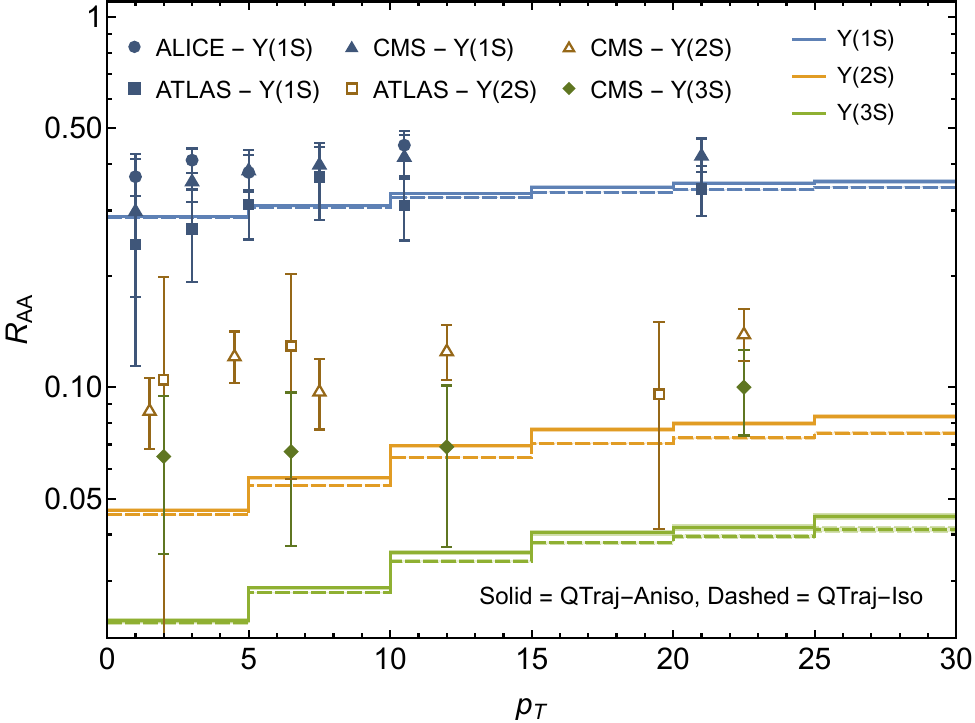}
	\end{center}
	\caption{Nuclear suppression factor, $R_{AA}$, of bottomonium s-wave states as a function of $N_\text{part}$ (left panel) and $p_T$ (right panel). The solid and dashed lines show the predictions of QTraj-Aniso and QTraj-Iso respectively. The experimental measurements shown are from the ALICE~\cite{ALICE:2020wwx}, ATLAS~\cite{ATLAS5TeV}, and CMS~\cite{Sirunyan:2018nsz,CMS-PAS-HIN-21-007} collaborations. Experimental error bars shown were obtained by adding statistical and systematic uncertainties in quadrature.}
	\label{fig01}
\end{figure}
%-------------------------
Fig.~\ref{fig01}, we present the nuclear suppression factor $R_{AA}$ of the $\Upsilon(1S)$, $\Upsilon(2S)$, and $\Upsilon(3S)$ states as a function of $N_{\mathrm{part}}$ (left panel) and transverse momentum $p_T$ (right panel), comparing the predictions of QTraj-Aniso and QTraj-Iso. The solid and dashed lines correspond to QTraj-Aniso and QTraj-Iso, respectively, while the experimental data from the ALICE~\cite{ALICE:2020wwx}, ATLAS~\cite{ATLAS5TeV}, and CMS~\cite{Sirunyan:2018nsz,CMS-PAS-HIN-21-007} collaborations are shown as circles, squares, and triangles. For the left panel, we apply a transverse momentum cut of $p_T < 30$ GeV. Both QTraj-Aniso and QTraj-Iso reproduce the expected sequential suppression pattern, $R_{AA}[\Upsilon(1S)] > R_{AA}[\Upsilon(2S)] > R_{AA}[\Upsilon(3S)]$, with the tightly bound ground state being least suppressed and the more weakly bound excited states showing stronger suppression. From the left panel, we observe that both frameworks provide a reasonable description of the $N_{\mathrm{part}}$ dependence of $R_{AA}[\Upsilon(1S)]$, with QTraj-Aniso showing a modest but systematic improvement in agreement with the experimental measurements, particularly in the mid-to-central collision region where medium effects are strongest. For the excited states, both approaches tend to predict somewhat stronger suppression for $\Upsilon(2S)$ than indicated by the data, although the anisotropic implementation again provides slightly better agreement than the isotropic baseline. The difference between QTraj-Aniso and QTraj-Iso is not dramatic, but it consistently indicates that momentum-space anisotropy improves the phenomenological description of bottomonium suppression. In the right panel, we present $R_{AA}[\Upsilon]$ as a function of $p_T$, where the results are averaged over centrality using the weight function $w(c)=\exp(-c/20)$ with $c\in[0,100]$. This weight function reflects the experimentally observed distribution of the number of $\Upsilon$ states versus centrality~\cite{Chatrchyan:2012np}. Both models predict a weak $p_T$ dependence of $R_{AA}$, with only a mild increase toward larger $p_T$. The slightly stronger suppression at low $p_T$ can be attributed to low-momentum wave packets spending, on average, a longer effective time inside the QGP fireball due to their smaller velocities. Here again, QTraj-Aniso lies systematically closer to the data than QTraj-Iso, indicating that anisotropic screening and dissociation effects improve the description of the in-medium evolution. A more complete treatment of the anisotropic complex potential, including higher-order corrections in the anisotropy parameter $\xi$ and a fully non-angle-averaged imaginary part, could further improve agreement with experimental observations.

%--------------------------
\begin{figure}[ht]
	\begin{center}
		\includegraphics[width=0.485\linewidth]{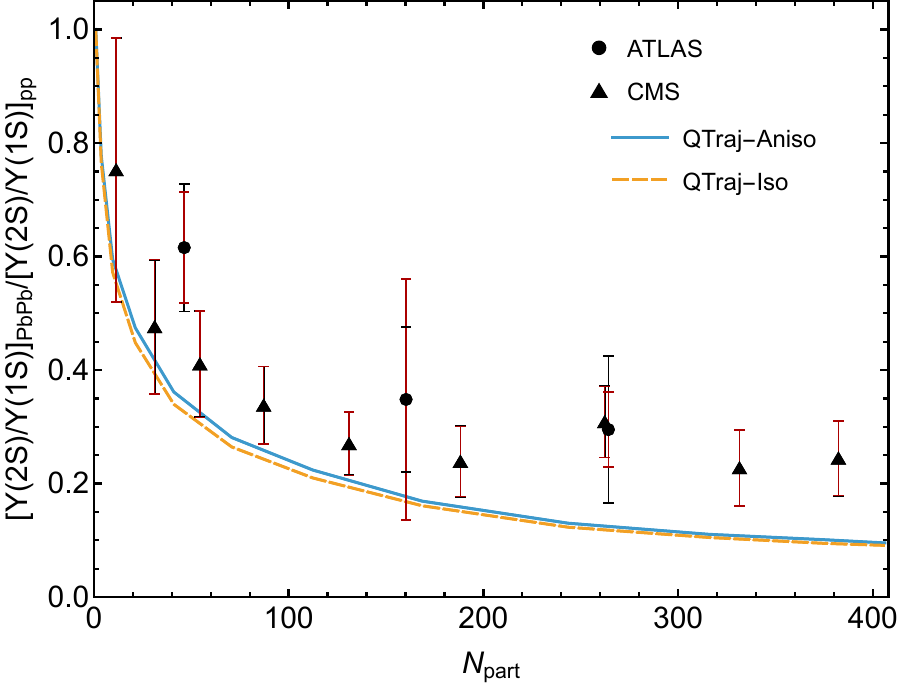}\hspace{2mm}
		\includegraphics[width=0.485\linewidth]{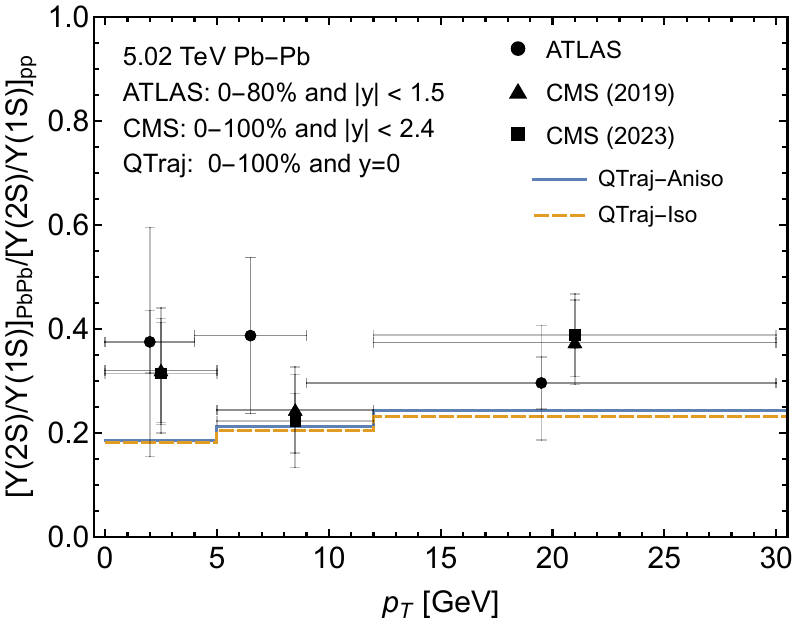}
	\end{center}
	\caption{Double ratio $[\Upsilon(2s)/\Upsilon(1s)]_\text{PbPb}/[\Upsilon(2s)/\Upsilon(1s)]_\text{pp}$ as a function of $N_\text{part}$ (left panel) and $p_T$ (right panel). The solid and dashed lines show the predictions of QTraj-Aniso and QTraj-Iso respectively. The experimental measurements shown are from the ATLAS~\cite{ATLAS5TeV} and CMS~\cite{Sirunyan:2018nsz, CMS-PAS-HIN-21-007} collaborations. }
	\label{fig02}
\end{figure}
%-------------------------
In Fig.~\ref{fig02}, we present the double ratio $[\Upsilon(2S)/\Upsilon(1S)]_{\mathrm{PbPb}}/[\Upsilon(2S)/\Upsilon(1S)]_{\mathrm{pp}}$ as a function of $N_{\mathrm{part}}$ (left panel) and transverse momentum $p_T$ (right panel), comparing the predictions of QTraj-Aniso and QTraj-Iso. We compare these predictions with data reported by the ATLAS and CMS collaborations in Refs.~\cite{ATLAS5TeV, Sirunyan:2018nsz, CMS-PAS-HIN-21-007}. For the ATLAS 2S to 1S double ratio, the black and red error bars correspond to statistical and systematic uncertainties, respectively. The data for the 2S to 1S double ratio labelled  `CMS (2023)' were inferred from the reported $p_T$-dependence of $R_{AA}[1S]$ and $R_{AA}[2S]$.  
For the ATLAS and `CMS (2019)' data, the collaborations reported their computed 2S to 1S double ratio directly. From the left panel, both QTraj-Aniso and QTraj-Iso reproduce the expected decrease of the double ratio with increasing $N_{\mathrm{part}}$, reflecting the stronger suppression of the less tightly bound $\Upsilon(2S)$ state in more central collisions. QTraj-Aniso shows a modest but systematic improvement over QTraj-Iso, particularly in the intermediate centrality region where the effects of medium anisotropy become more relevant. In the right panel, we show the $p_T$ dependence of the same double ratio. Both models predict a weak dependence on transverse momentum, with only a mild increase toward larger $p_T$, consistent with the behavior already observed for the individual nuclear modification factors. Here again, the anisotropic implementation lies slightly closer to the experimental measurements than the isotropic baseline, indicating that anisotropic screening and dissociation effects improve the relative description of excited-state suppression.

%--------------------------
\begin{figure}[ht]
	\begin{center}
		\includegraphics[width=0.485\linewidth]{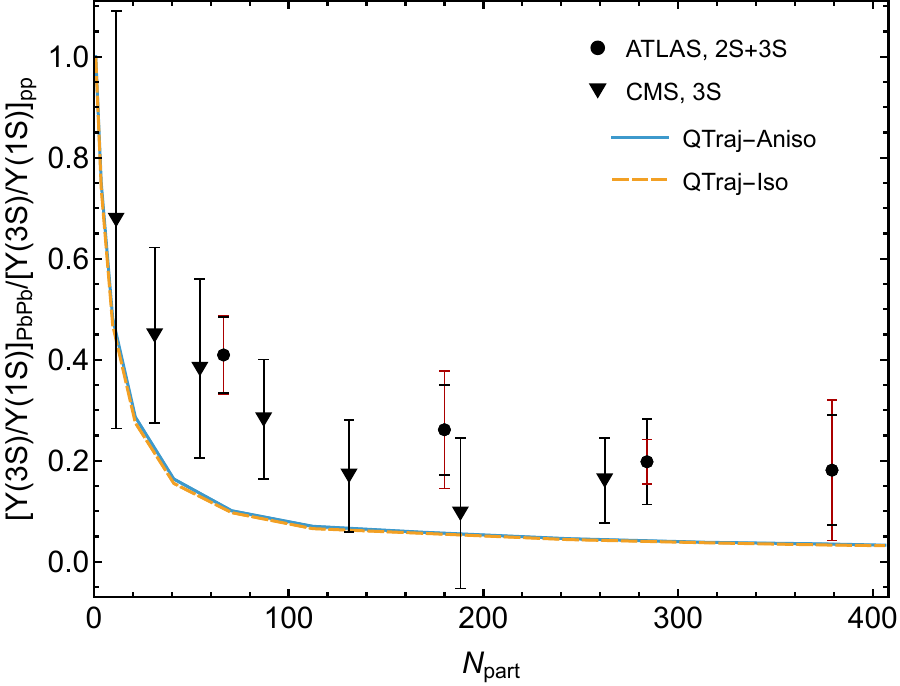}
	\end{center}
	\caption{Double ratio $[\Upsilon(3S)/\Upsilon(1S)]_\text{PbPb}/[\Upsilon(3S)/\Upsilon(1S)]_\text{pp}$ as a function of $N_\text{part}$. Line styles and experimental data sources are the same as Fig.~\ref{fig02}.}
	\label{fig03}
\end{figure}
\begin{figure}[ht]
	\begin{center}
		\includegraphics[width=0.485\linewidth]{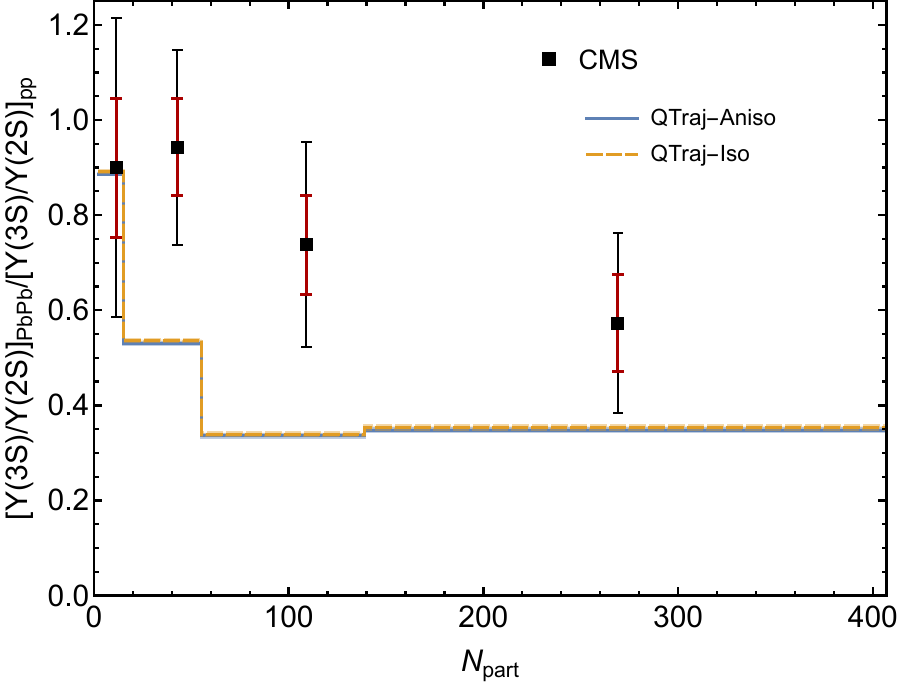}\hspace{2mm}
		\includegraphics[width=0.485\linewidth]{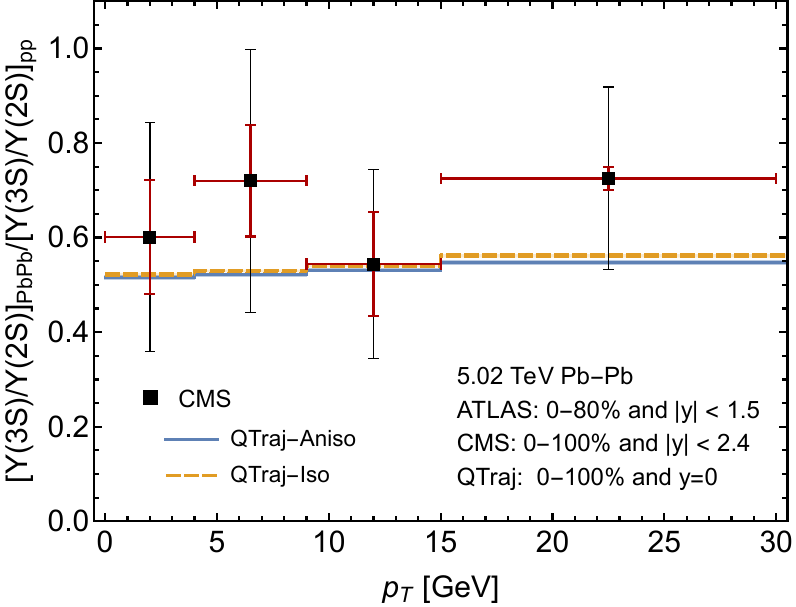}
	\end{center}
	\caption{Double ratio $[\Upsilon(3S)/\Upsilon(2S)]_\text{PbPb}/[\Upsilon(3S)/\Upsilon(2S)]_\text{pp}$ as a function of $N_\text{part}$ (left panel) and $p_T$ (right panel). Line styles are the same as Fig.~\ref{fig02}.  The centrality classes used were 0-30\%, 30-50\%, 50-70\%, and 70-90\%.   Experimental data are from Ref.~\cite{CMS-PAS-HIN-21-007}.}
	\label{fig04}
\end{figure}

%-------------------------
In Fig.~\ref{fig03}, we present the double ratio $[\Upsilon(3S)/\Upsilon(1S)]_{\mathrm{PbPb}}/[\Upsilon(3S)/\Upsilon(1S)]_{\mathrm{pp}}$ as a function of $N_{\mathrm{part}}$, comparing the predictions of QTraj-Aniso and QTraj-Iso.  In the case of the CMS 3S to 1S double ratio, we inferred this observable from the 3S to 2S double ratio reported by CMS (shown in Fig.~\ref{fig04}) and the inferred 2S to 1S double ratio presented in Fig.~\ref{fig02}.  
For the ATLAS 3S to 1S double ratio, we use their reported combined 2S+3S to 1S double ratio~\cite{Sirunyan:2018nsz}. As expected, both QTraj-Aniso and QTraj-Iso predict a strong decrease of the double ratio with increasing $N_{\mathrm{part}}$, reflecting the rapid melting of the loosely bound $\Upsilon(3S)$ state in central collisions. Nevertheless, the predicted double ratio remains somewhat smaller than the experimental measurements, indicating that both QTraj-Aniso and QTraj-Iso still slightly overestimate the suppression of the $\Upsilon(3S)$ state. Although the improvement from the anisotropic treatment is not dramatic, it consistently shifts the prediction in the correct direction.

%--------------------------
%-------------------------
In Fig.~\ref{fig04}, we present the double ratio $[\Upsilon(3S)/\Upsilon(2S)]_{\mathrm{PbPb}}/[\Upsilon(3S)/\Upsilon(2S)]_{\mathrm{pp}}$ as a function of $N_{\mathrm{part}}$ (left panel) and transverse momentum $p_T$ (right panel), comparing the predictions of QTraj-Aniso and QTraj-Iso. We compare our prediction with data reported by the CMS collaborations in Ref.~\cite{CMS-PAS-HIN-21-007}. From the left panel, both QTraj-Aniso and QTraj-Iso predict a decreasing trend of the double ratio with increasing $N_{\mathrm{part}}$, reflecting the stronger suppression of the more weakly bound $\Upsilon(3S)$ state compared to $\Upsilon(2S)$ as the medium becomes hotter and denser in central collisions. The anisotropic and isotropic predictions are very close to each other for this observable, indicating that the relative suppression between these two excited states is less sensitive to medium anisotropy than the comparison with the ground state shown in Figs.~\ref{fig02} and \ref{fig03}. In the right panel, we show the $p_T$ dependence of the same double ratio. Both models predict a very weak dependence on transverse momentum, with an approximately flat behavior across the measured $p_T$ range. This is consistent with the expectation that both excited states experience similarly strong in-medium suppression and therefore their ratio exhibits only a mild momentum dependence. As can be seen from both panels, both QTraj-Aniso and QTraj-Iso underpredict the measured double ratio, indicating that the present framework still overestimates the suppression of the $\Upsilon(3S)$ state relative to $\Upsilon(2S)$.

%--------------------------
\begin{figure}[ht]
	\begin{center}
		\includegraphics[width=0.95\linewidth]{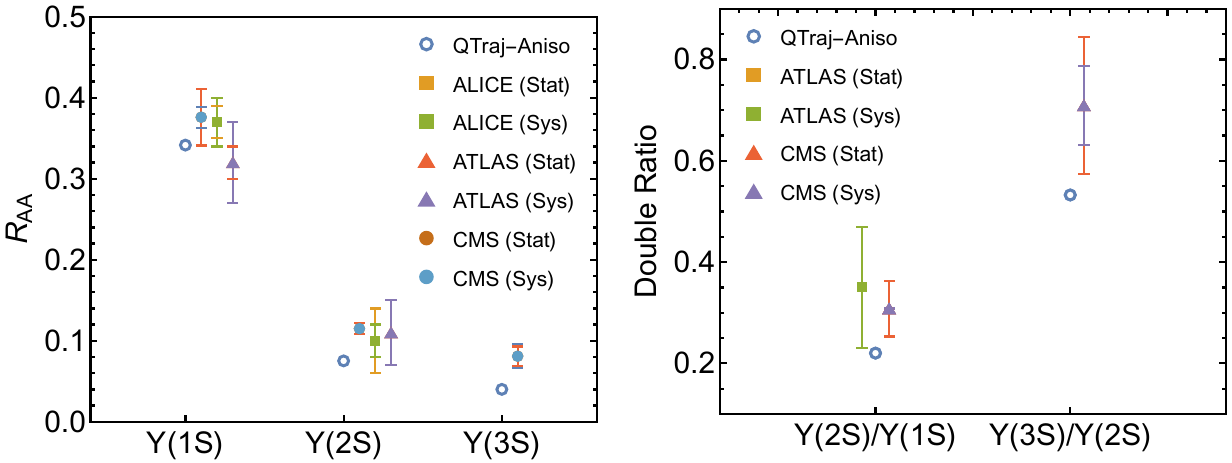}
	\end{center}
	\caption{Integrated nuclear suppression factors $R_{AA}$ (left panel) and double ratios (right panel) for bottomonium s-wave states. The open blue circles (QTraj-Aniso) represent theoretical calculations, while the filled markers show experimental results from ALICE~\cite{ALICE:2020wwx}, ATLAS~\cite{ATLAS5TeV}, and CMS~\cite{Sirunyan:2018nsz, CMS-PAS-HIN-21-007} collaborations with statistical and systematic uncertainties indicated separately.}
	\label{fig05}
\end{figure}
%-------------------------

In Figure \ref{fig05}, we present integrated nuclear suppression factors  $R_{AA}$ (left panel) and double ratios (right panel) for bottomonium s-wave states in Pb–Pb collisions at the LHC. The open blue circles (QTraj-Aniso) represent theoretical calculations, while the filled markers show experimental results from ALICE~\cite{ALICE:2020wwx}, ATLAS~\cite{ATLAS5TeV}, and CMS~\cite{Sirunyan:2018nsz, CMS-PAS-HIN-21-007} collaborations with statistical and systematic uncertainties indicated separately. The values are integrated over transverse momentum up to 40 GeV and over the full rapidity range, yielding kinematics-averaged suppression factors suitable for direct comparison with experimental measurements. In the left panel, the sequential suppression pattern is evident, with the  $\Upsilon(1S)$ being least suppressed and the excited states  $\Upsilon(2S)$,  $\Upsilon(3S)$ exhibiting stronger suppression, consistent with both the QTraj-Aniso framework and LHC data. In the right panel,  the integrated double ratios $R_{AA}(\Upsilon(2S))/R_{AA}(\Upsilon(1S))$ and $R_{AA}(\Upsilon(3S))/R_{AA}(\Upsilon(2S))$ are presented. These ratios highlight the relative suppression between excited and ground states, providing a cleaner probe of in-medium effects by reducing common systematic uncertainties. The theoretical QTraj-Aniso results (open blue circles) are compared against ATLAS and CMS data, with both experiments showing consistency with the expected sequential melting picture. Overall, the comparison indicates that the QTraj-Aniso framework captures the qualitative hierarchy of bottomonium suppression and provides reasonable agreement with the experimental measurements across different collision systems.

%%%%%%%%%%%%%%%%%%%%%%%%%%%%%%%%%%%%%%%%%%%%%%%%%%%%%%%%%%%%%%
\subsection{Elliptic flow~\texorpdfstring{$v_{2}$}{v2}}
\label{sec:v2}
%%%%%%%%%%%%%%%%%%%%%%%%%%%%%%%%%%%%%%%%%%%%%%%%%%%%%%%%%%%%%%

%--------------------------
\begin{figure}[ht]
	\begin{center}
		\includegraphics[width=0.475\linewidth]{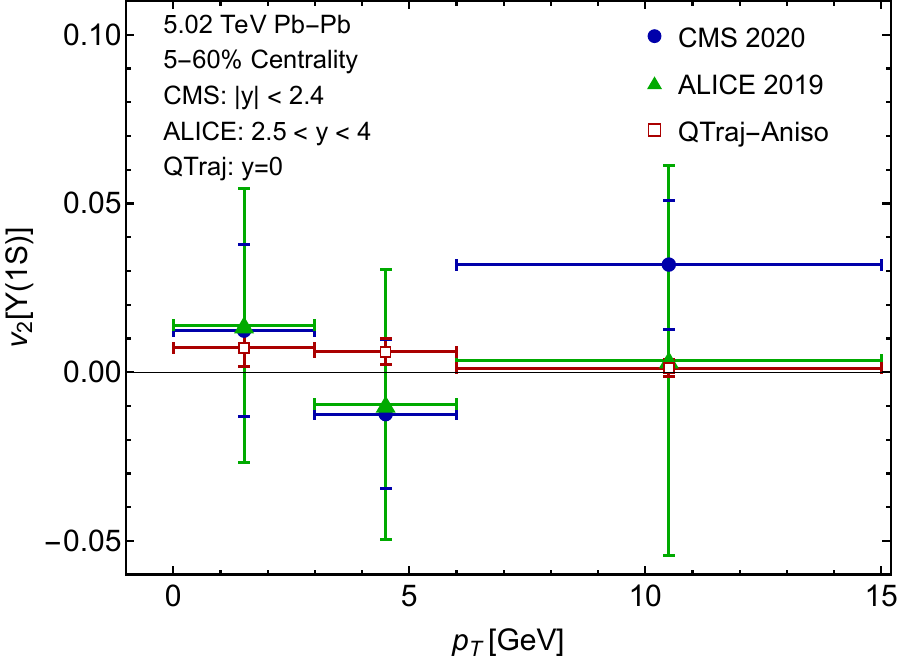}\hspace{2mm}
		\includegraphics[width=0.495\linewidth]{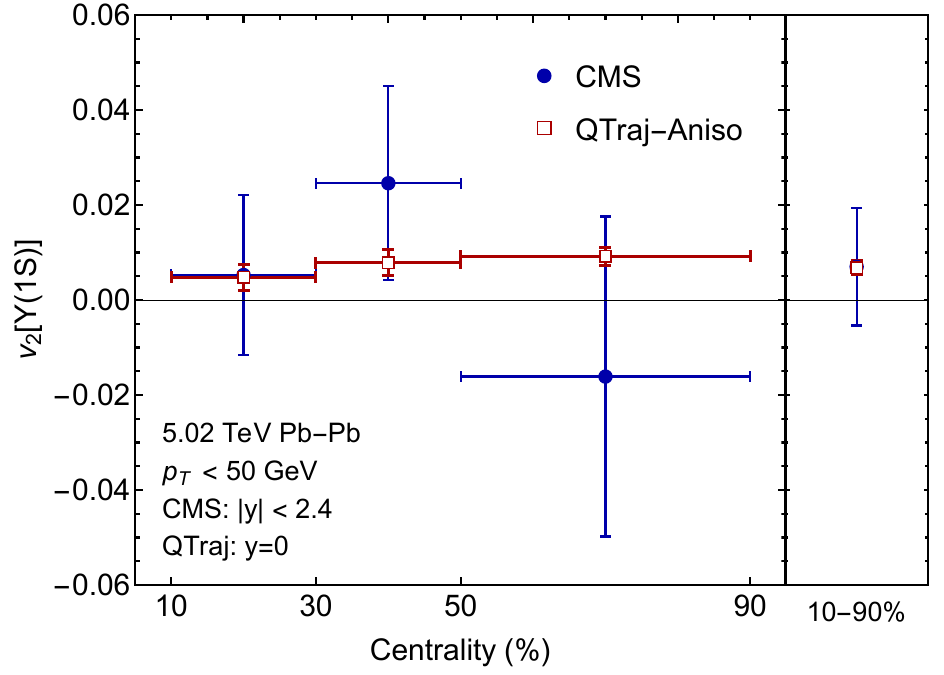}
	\end{center}
	\caption{The elliptic flow $v_2[\Upsilon(1S)]$ as a function of $p_T$ in three $p_T$-bins (left panel).Centrality dependence of $v_2[\Upsilon(1S)]$ shown in 10-30\%, 30-50\%, 50-90\%, and 10-90\% centrality bins (right panel). Open squares are predictions of QTraj-Aniso, and the data are from the ALICE \cite{Acharya:2019hlv} and CMS \cite{CMS:2020efs} collaborations.}
	\label{fig06}
\end{figure}
%-------------------------
A comparison of QTraj-Aniso for $v_2[\Upsilon(1S)]$ with experimental data collected by the ALICE \cite{Acharya:2019hlv} and CMS \cite{CMS:2020efs} collaborations in three different transverse momentum bins, 0-4, 4-6, and 6-15 GeV, is shown in the left panel of Fig.~\ref{fig06}. The QTraj-Aniso predictions and experimental results are integrated over centrality in the range \mbox{5-60\%}. One sees from this figure that QTraj-Aniso predicts results consistent with $v_2$ being nearly zero, slightly negative, and a small but positive value in the lowest $p_T$ bin, central $p_T$ bin, and highest $p_T$ bin, respectively. A similar trend has also been predicted earlier by an adiabatic approximation based model in Ref.~\cite{Bhaduri:2020lur} where the negative $v_2$ observed in the central $p_T$ bin is related to the transverse expansion of the QGP overtaking bottomonia states which have escaped from near the surface of the QGP. Returning to the theory to data comparison, Fig.~\ref{fig06} demonstrates that QTraj-Aniso has a reasonable agreement with the available experimental data given current experimental uncertainties.   Finally, we note that even given the experimental uncertainties, we see a similar trend in the experimental results as a function of $p_T$ as predicted by QTraj-Aniso. In the right panel of Fig.~\ref{fig06}, we present the centrality dependence of $v_2[\Upsilon(1S)]$ in 10-30\%, 30-50\%, 50-90\%, and 10-90\% centrality bins and compare our QTraj-Aniso predictions with experimental data from the CMS collaboration~\cite{CMS:2020efs}. We show a comparison of the QTraj-Aniso prediction with the experimental result integrated over 10-90\% centrality in the right side of the right panel. As can be seen from this figure, in the case of the integrated $v_2[\Upsilon(1S)]$ in the 10-90\% bin, there is quite reasonable agreement, within experimental uncertainties, between the QTraj-Aniso prediction and the data reported by the CMS collaboration. In the left side of the right panel of the Fig.~\ref{fig06}, good agreement between QTraj-Aniso predictions and CMS data can be seen in the 10-30\% bin; however, there are larger differences in the other two centrality bins, but remain within approximately $2\sigma$ of the experimental data. Hopefully, higher experimental statistics from future runs will reduce the experimental uncertainties in the near future.

%--------------------------
\begin{figure}[ht]
	\begin{center}
		\includegraphics[width=1.05\linewidth]{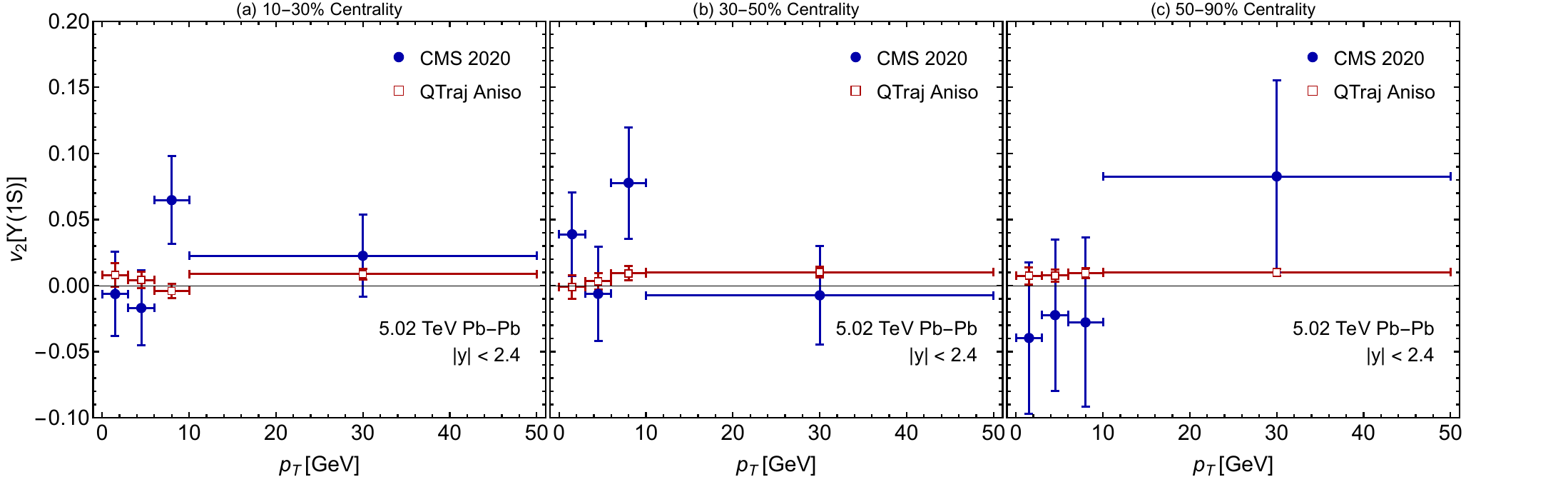}
	\end{center}
	\caption{Comparison of the elliptic flow coefficient $v_{2}$ of $\Upsilon(1S)$ as a function of transverse momentum $p_{T}$ for three centrality classes: (a) 10--30\%, (b) 30--50\%, and (c) 50--90\%.     The results from the QTraj-Aniso (red open squares) 
    are compared with CMS 2020 data (blue circles)~\cite{CMS:2020efs}. }
	\label{fig07}
\end{figure}
%-------------------------
Figure~\ref{fig07} shows the elliptic flow coefficient $v_{2}$ of 
$\Upsilon(1S)$ state as a function of transverse momentum $p_{T}$ for three different centrality intervals: 10--30\%, 30--50\%, and 50--90\%. The results from our QTraj-Aniso are compared with CMS measurements~\cite{CMS:2020efs}. Overall, both the model and the experimental data indicate small values of $v_{2}$, consistent with the weak collective flow of bottomonia. The QTraj-Aniso results predict a nearly flat $p_{T}$ dependence of $v_{2}$, remaining close to zero across the considered centrality classes. While the experimental uncertainties are large, the CMS data also favor values compatible with zero within errors, particularly at higher $p_{T}$. Some deviations are observed at low $p_{T}$, where the data 
suggest a slightly positive $v_{2}$ in semi-central collisions, although still consistent with the model within uncertainties.

%--------------------------
\begin{figure}[ht]
	\begin{center}
		\includegraphics[width=0.485\linewidth]{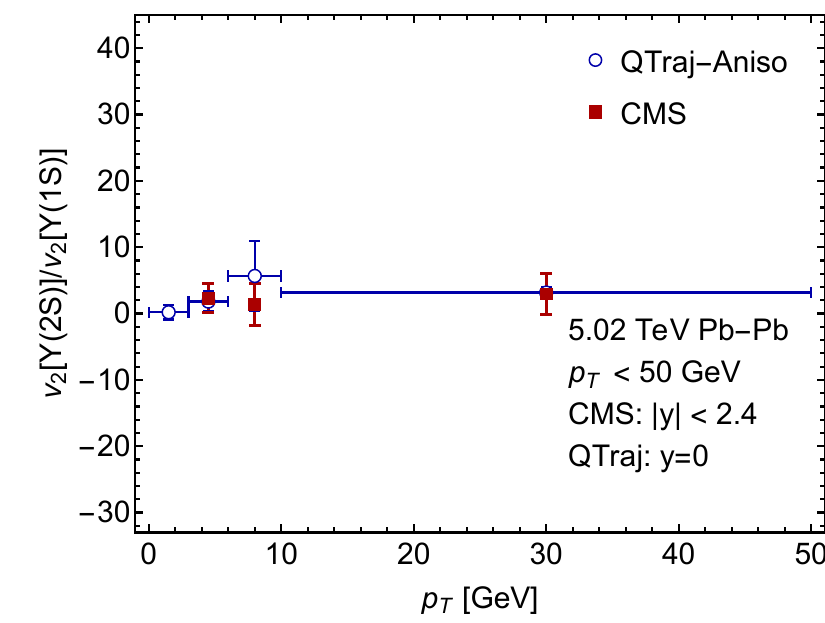}
	\end{center}
	\caption{The elliptic flow ratio $v_2[\Upsilon(2S)]$/$v_2[\Upsilon(1S)]$ as a function of $p_T$. The results from the QTraj-Aniso (blue open circle) are compared with CMS 2020 data (red square)~\cite{CMS:2020efs}.}
	\label{fig08}
\end{figure}
%-------------------------
Figure~\ref{fig08} presents the ratio of elliptic flow coefficients 
$v_{2}[\Upsilon(2S)]/v_{2}[\Upsilon(1S)]$ as a function of $p_T$. The results from the 
QTraj-Aniso calculations are compared with CMS measurements~\cite{CMS:2020efs}. Within the experimental uncertainties, the ratio is consistent with unity across the measured $p_T$ range, indicating no significant relative enhancement of the elliptic flow between the excited and ground states. The QTraj-Aniso results predict a nearly flat ratio, also consistent with the CMS data, suggesting that both $\Upsilon(1S)$ and $\Upsilon(2S)$ experience a similarly weak coupling to the collective medium flow. This supports the picture that bottomonium elliptic flow is generally small and that differences between states are difficult to resolve with the current experimental precision.

%--------------------------
\begin{figure}[ht]
	\begin{center}
		\includegraphics[width=0.40\linewidth]{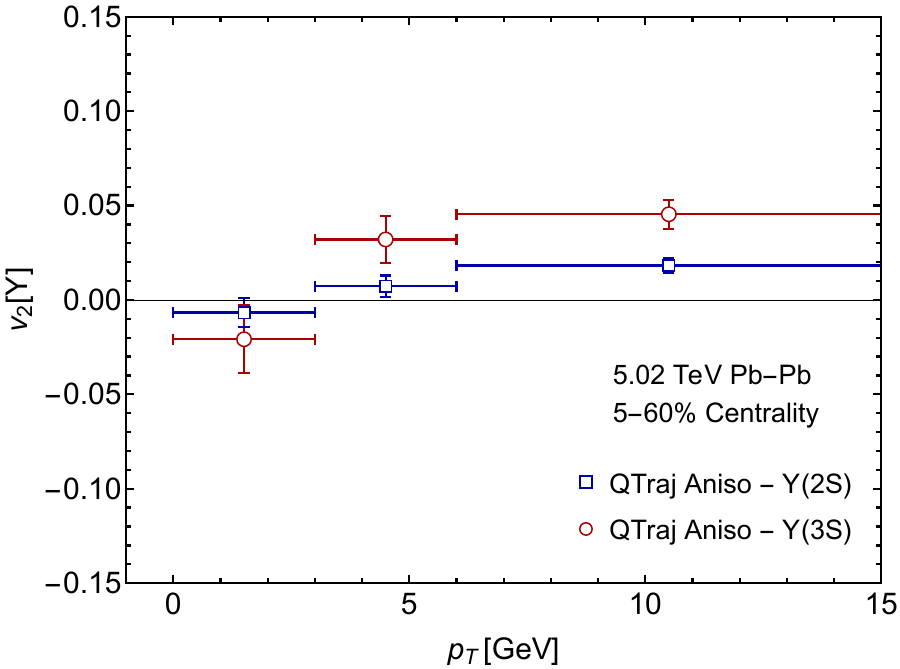}\hspace{2mm}
		\includegraphics[width=0.55\linewidth]{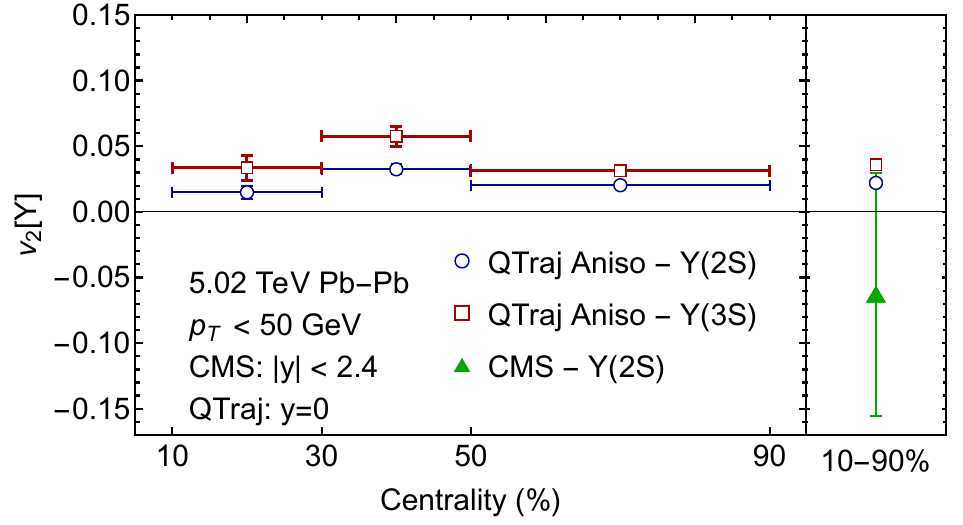}
	\end{center}
	\caption{The elliptic flow $v_2[\Upsilon(2S)]$, and $v_2[\Upsilon(3S)]$  as a function of $p_T$ (left panel).Centrality dependence of $v_2[\Upsilon(2S)]$, and $v_2[\Upsilon(3S)]$ shown in 10-30\%, 30-50\%, 50-90\%, and 10-90\% centrality bins (right panel).  In the 10-90\% class, we include recent data reported by the CMS collaboration for integrated $v_2[\Upsilon(2S)]$ \cite{CMS:2020efs}. }
	\label{fig09}
\end{figure}
%-------------------------
In the left panel of Fig.~\ref{fig09}, we present QTraj-Aniso predictions for the elliptic flow of $\Upsilon(2S)$ and $\Upsilon(3S)$ states as a function of transverse momentum in the 5-60\% centrality bin. As can be seen from this figure, a sizable $v_2$ is predicted by QTraj-Aniso for the $\Upsilon(3S)$ because of the strong path length dependence of $R_{AA}[\Upsilon(3S)]$ between the short and long sides of the QGP fireball. Similar to the $\Upsilon(1S)$, QTraj-Aniso predicts a negative $v_2$ for the $\Upsilon(2S)$ in the lowest two $p_T$-bins. The reason behind this negative $v_2$ is again that the transverse expansion of the QGP occurs more rapidly along the short side than the long side which has the affect of overtaking bottomonium states which had previously escaped the QGP with $\phi \sim 0$. Although the sign of $v_2$ becomes negative in the lowest $p_T$ region, the magnitude of the elliptic flow still reflects the expected sequential ordering, with the more weakly bound $\Upsilon(3S)$ state showing the strongest response. In the highest $p_T$ bin, both excited states exhibit positive elliptic flow.

In the right panel of Fig.~\ref{fig09}, we show the centrality dependence of $v_2[\Upsilon(2S)]$ and $v_2[\Upsilon(3S)]$ in the intervals 10--30\%, 30--50\%, 50--90\%, and the integrated 10--90\% bin. One centrality-integrated data point for $v_2[\Upsilon(2S)]$ from the CMS collaboration is shown as a green triangle in the 10-90\% panel. As can be seen from the right side of the figure, the integrated result for $v_2[\Upsilon(2S)]$ falls within the experimental uncertainties, albeit at the very top end of them. We again hope that more statistics will help in making more constraining comparisons in the near future.  Note that for  $v_2[\Upsilon(2S)]$ and $v_2[\Upsilon(3S)]$ one sees a larger systematic uncertainty than for $v_2[\Upsilon(1S)]$.  Given these larger uncertainties, however, one still observes an ordering of the elliptic flow of the excited states with $v_2[\Upsilon(3S)] > v_2[\Upsilon(2S)]$ in all centrality bins.

%%%%%%%%%%%%%%%%%%%%%%%%%%%%%%%%%%%%%%%%%%%%%%%%%%%%%%%%%%%%%%
\section{Conclusions and outlook}
\label{sec09}
%%%%%%%%%%%%%%%%%%%%%%%%%%%%%%%%%%%%%%%%%%%%%%%%%%%%%%%%%%%%%%
In this work, we have presented a comprehensive framework for studying the in-medium dynamics of bottomonium states in a quark-gluon plasma (QGP) using the quantum trajectories (QTraj) method. Our approach incorporates both the isotropic Karsch-Mehr-Satz (KMS) potential and its momentum-space anisotropic generalization, allowing for a systematic comparison between isotropic and anisotropic medium effects. The real part of the anisotropic potential, $V_R^{\rm aniso}$, was constructed via a minimal extension using an anisotropic screening mass. For the imaginary part, $V_I^{\rm aniso}$, we derived the angle-averaged leading-order behavior in the anisotropy parameter $\xi$ and constructed a phenomenological model that interpolates smoothly between the correct asymptotic limits, $\sim (1-\xi/6)$ for small $\xi$ and $\sim 1/\sqrt{\xi}$ for large $\xi$. By solving the real-time Schr\"{o}dinger equation with this anisotropic complex potential on a realistic 3+1D hydrodynamic background, we computed the nuclear modification factors $R_{AA}$, double ratios, and elliptic flow coefficients $v_2$ for the $\Upsilon(1S)$, $\Upsilon(2S)$, and $\Upsilon(3S)$ states in Pb-Pb collisions at $\sqrt{s_{NN}}=5.02$ TeV, including the feed-down effect from higher states.

Both QTraj-Iso and QTraj-Aniso reproduce the characteristic sequential suppression pattern, $R_{AA}(\Upsilon(1S)) > R_{AA}(\Upsilon(2S)) > R_{AA}(\Upsilon(3S))$, reflecting the hierarchy of binding energies and in-medium dissociation rates. The tightly bound ground state exhibits the weakest suppression, while the more weakly bound excited states are increasingly suppressed with collision centrality. For the ground state, both frameworks provide a reasonable description of the measured $R_{AA}$, with QTraj-Aniso showing a modest but systematic improvement over the isotropic baseline, particularly in the mid-central and central collision regions where medium effects are strongest. Similar behavior is observed for the $p_T$ dependence and for the $\Upsilon(2S)/\Upsilon(1S)$ and $\Upsilon(3S)/\Upsilon(1S)$ double ratios, where the anisotropic treatment consistently moves the predictions closer to the experimental data.

A notable discrepancy remains for the $\Upsilon(2S)$ and $\Upsilon(3S)$ excited states, where both QTraj-Iso and QTraj-Aniso still predict somewhat stronger suppression than indicated by the experimental double ratios. This suggests that the present framework may overestimate the dissociation of weakly bound states, likely due to the simplified leading-order and angle-averaged treatment of the imaginary part of the anisotropic potential. Nevertheless, the anisotropic implementation consistently improves the phenomenological description relative to the isotropic case, demonstrating that momentum-space anisotropy plays a non-negligible role even for suppression observables.

An important result of this work is the prediction of nonzero elliptic flow $v_2$ for bottomonium states. In contrast to the suppression observables, where anisotropy provides a moderate quantitative improvement, the description of $v_2$ is intrinsically tied to anisotropic medium effects. QTraj-Aniso naturally produces a small but finite $v_2$ for $\Upsilon(1S)$, with a nontrivial dependence on transverse momentum and centrality that is in reasonable agreement with ALICE and CMS measurements within current uncertainties. The slightly negative $v_2$ observed in intermediate and low-$p_T$ bins can be understood from the competition between path-length dependent suppression, escape probability, and transverse medium expansion.

For the excited states, the magnitude of the elliptic flow follows the expected sequential pattern, $v_2(\Upsilon(3S)) > v_2(\Upsilon(2S)) > v_2(\Upsilon(1S))$,  with the more weakly bound states exhibiting a stronger anisotropic response. In particular, $\Upsilon(3S)$ shows the largest elliptic flow, followed by $\Upsilon(2S)$ and $\Upsilon(1S)$, although the sign of $v_2$ can become negative in the lowest $p_T$ region due to transverse expansion effects. This behavior further supports the interpretation that the elliptic flow of quarkonium is a sensitive probe of both path-length dependent suppression and the dynamical anisotropy of the QGP medium.

Several directions for future work follow naturally from these results. A more complete first-principles calculation of the imaginary part of the anisotropic potential, beyond the present leading-order and angle-averaged approximation, is expected to improve the description of the $\Upsilon(2S)$ and $\Upsilon(3S)$ states. Including higher-order corrections in the anisotropy parameter $\xi$ and a fully non-angle-averaged treatment of the complex potential would allow for a more quantitative understanding of the dissociation dynamics. Event-by-event initial conditions and full viscous hydrodynamic evolution could further improve the description of fluctuating medium anisotropies and their impact on $v_2$. While regeneration is expected to be subdominant for bottomonium at LHC energies, a more systematic assessment of its contribution, particularly for $\Upsilon(1S)$ at very low $p_T$, would also be valuable.

Future high-statistics measurements of the elliptic flow of $\Upsilon(2S)$ and $\Upsilon(3S)$, together with improved double-ratio measurements, will provide particularly stringent tests of the present framework. These observables directly probe the interplay between binding energy, dissociation, and path-length dependence, and therefore offer strong constraints on the in-medium heavy-quark potential. Extending the present formalism to other collision systems, including p--Pb, high-multiplicity pp, and future Electron-Ion Collider (EIC) environments, will provide further opportunities to test its robustness.

In conclusion, this work establishes a clear connection between momentum-space anisotropy of the QGP and key bottomonium observables. The isotropic framework provides an important baseline, while the anisotropic generalization yields systematic improvements for suppression observables and is essential for generating realistic elliptic flow. These results highlight the quantum trajectories method as a powerful tool for performing real-time tomography of the hottest matter created in the laboratory, providing new insight into the transport properties and dynamical evolution of the quark-gluon plasma.
%%%%%%%%%%%%%%%%%%%%%%%%%%%%%
\section*{Acknowledgments}
%%%%%%%%%%%%%%%%%%%%%%%%%%%%%
A.I. wishes to thank Michael Strickland for valuable discussions and constructive suggestions during the initial phase of this project. A.I. also acknowledges the support and hospitality of the Institute of Particle Physics at Central China Normal University during the postdoctoral appointment. Numerical computations were performed on the high-performance computing facilities of the Nuclear Science Computing Center at Central China Normal University (NSC$^3$).
%%%%%%%%%%%%%%%%%%%%%%%%%%%%%%%%%%%%%%%%%%%%%%%%%
\appendix
%%%%%%%%%%%%%%%%%%%
\section*{Appendix}
%%%%%%%%%%%%%%%%%%%
\section{Derivation of the~\texorpdfstring{$f(\xi)$}{f(xi)}~-~function}
\label{app:a}
%%%%%%%%%%%%%%%%%%%%%%%%%%%%%%%%%%%%%%%%%%%%%%%%
In anisotropic hydrodynamics (aHydro), the evolution equation for a general moment can be written as \cite{Strickland:2018ayk}
\be
{\cal M}^{nm}_{\rm aHydro}(\tau) = \frac{\Gamma(n+2m+2) \Lambda^{n+2m+2}(\tau)}{(2\pi)^2} {\cal H}^{nm}\!\left( \alpha(\tau) \right) ,
\label{mres_ahydro}
\ee
 where in case of Bjorken flow, the Minkowski-space components of the four-flow are \mbox{$u^\mu = (t/\tau,0,0,z/\tau)$}, with $\tau = \sqrt{t^2-z^2}$ is the longitudinal proper-time.
 
 Here,
 \begin{equation}
 \alpha(\tau) = 1/\sqrt{1+\xi(\tau)}\, ,
 \end{equation}
and
\be
{\cal H}^{nm}(y) = \tfrac{2y^{2m+1}}{2m+1}  {}_2F_1(\tfrac{1}{2}+m,\tfrac{1-n}{2};\tfrac{3}{2}+m;1-y^2)  \, ,
\ee
with ${}_2F_1$ being hypergeometric function.

In equilibrium, we have  
\be
{\cal M}^{nm}_{\rm eq}(\tau) = \frac{\Gamma(n+2m+2) T^{n+2m+2}(\tau)}{2\pi^2 (2m +1)} \, .
\label{meqres2}
\ee
Now the scaled moments is given by \cite{Strickland:2018ayk}
\begin{eqnarray}
\overline{\cal M}^{nm}_{\rm aHydro}(\tau) &=& \frac{{\cal M}^{nm}_{\rm aHydro}(\tau)}{{\cal M}^{nm}_{\rm eq}(\tau)}\nonumber\\
&=& 2^{(n+2m-2)/4} (2m+1) \frac{{\cal H}^{nm}(\alpha)}{[{\cal H}^{20}(\alpha)]^{(n+2m+2)/4}}\, . \label{eq:ahydromoms}
\end{eqnarray}
We also have the isotropic Debye mass \cite{Strickland:2011aa}
\begin{equation}
m^2_D=-\frac{g^2}{2\pi^2}\int_{0}^{\infty}dp~p^2\frac{df_{\mathrm{iso}}}{dp}\,\, ,
\end{equation}
where $p^2 \equiv \mathbf{p}^2=p^2_\perp+p^2_z$. Now integrating by parts
\begin{equation}
\int UV dx = U \int V dx - \int \left[\frac{dU}{dx}\int V dx\right]dx\,\, ,
\end{equation}
with $U = p^2 $, $V = \frac{df_{\mathrm{iso}}}{dp} $ ; we get 
\begin{eqnarray}
m^2_D &=& -\frac{g^2}{2\pi^2} \left[p^2 f_{\mathrm{iso}}\right]_{0}^{\infty} + \frac{g^2}{\pi^2}\int_{0}^{\infty}dp~pf_{\mathrm{iso}}\nonumber\\
&=& 0 + \frac{g^2}{\pi^2}\int_{0}^{\infty}dp~pf_{\mathrm{iso}}\nonumber\\
\Rightarrow m^2_D &=& \frac{g^2}{\pi^2}\int_{0}^{\infty}dp~pf_{\mathrm{iso}}\,\, .
\end{eqnarray}
Now we can write an expression for the anisotropic screen scale or anisotropic Debye mass as 
\begin{equation}
\mu^2 = m^2_{D,~ \mathrm{aniso}} = \frac{g^2}{\pi^2}\int_{0}^{\infty}dp~pf(p)\label{A.9}
\end{equation}
From Ref. \cite{Strickland:2018ayk}, we can also write down the evolution equation for a general moment in anisotropic hydrodynamics (aHydro) as 
\begin{eqnarray}
 \mathcal{M} ^{nm}_{\rm aHydro} &=& \int dP~(p.u)^n (p.z)^{2m} f(p)\nonumber\\
 \Rightarrow \mathcal{M} ^{00}_{\rm aHydro} &=& \int dP~f(p)\label{A.10}\\
 &=& \frac{1}{(2\pi)^3}\int d\Omega \int_{0}^{\infty} \frac{dp~p^2}{p}f(p)\label{A.11}\\
 &=& \frac{1}{(2\pi)^3}~ 4\pi \int_{0}^{\infty} \frac{dp~p^2}{p}f(p)\label{A.12}\\
 &=& \frac{1}{2\pi^2}\int_{0}^{\infty}dp~pf(p)\,\,.\label{A.13}
\end{eqnarray}
where in going from Eq. \ref{A.10} to Eq. \ref{A.11}, we use $dP = \frac{d^3p}{(2\pi)^3}\frac{1}{p}$ and $d^3p = d\Omega~dp~p^2$ , in Eq. \ref{A.12} we use $\int d\Omega = 4\pi$ .

Inserting Eq. \ref{A.13} in Eq. \ref{A.9}, we get 
\begin{equation}
\mu^2 = 2 g^2 \mathcal{M} ^{00}_{\rm aHydro} \, \, .
\end{equation}
Similarly, we can write 
\begin{equation}
m^2_D = m^2_{D,~ \mathrm{eq}} = 2 g^2 \mathcal{M} ^{00}_{\rm eq}\,\, .
\end{equation}
Hence,
\begin{equation}
\frac{\mu^2}{m^2_D} = \frac{\mathcal{M} ^{00}_{\rm aHydro}}{\mathcal{M} ^{00}_{\rm eq}} = \overline{\mathcal{M}}^{00}_{\rm aHydro} \,\, . \label{A.16}
\end{equation}

With the help of Eq. \ref{A.16} and Eq. \ref{eq:ahydromoms}, we extract the $f(\xi)$-function as follows
\begin{eqnarray}
f(\xi) &=& \frac{\mu^2}{m^2_D}\nonumber\\
&=&\overline{\cal M}^{00}_{\rm aHydro}(\tau)\nonumber\\
&=&\overline{\cal M}^{00}_{\rm aHydro}(\xi)\nonumber\\
&=&\frac{\sqrt{2}(1+\xi)\arcsin{\left(\sqrt{\frac{\xi}{1+\xi}}\right)}}{\sqrt{\xi}\sqrt{1+\xi+\frac{(1+\xi)^2 \arcsin{\left(\sqrt{\frac{\xi}{1+\xi}}\right)}}{\sqrt{\xi}}}}\, ,
\end{eqnarray}
which is Eq. \ref{ffunc}.

%%%%%%%%%%%%%%%%%%%
%\section{Extra plots}
%\label{app:b}
%%%%%%%%%%%%%%%%%%%%%%%%%%%%%%%%%%%%%%%%%%%%%%%%

%--------------------------
%\begin{figure}[ht]
	%\begin{center}
		%\includegraphics[width=1\linewidth]{figures/16-v2vscentralityin20perbin.pdf}
	%\end{center}
	%\caption{Elliptic flow coefficient $v_2$ of bottomonium states as a function of centrality in 20\% wide bins. The bars represent the extracted $v_2$ values with statistical uncertainties shown as error bars. The results are consistent with vanishing flow in central and peripheral collisions, while a modest positive $v_2$ emerges in mid-central events, reflecting the expected collective anisotropy of the quark--gluon plasma medium.}
	%\label{fig10}
%\end{figure}
%-------------------------

%--------------------------
%\begin{figure}[ht]
	%\begin{center}
		%\includegraphics[width=0.8\linewidth]{figures/15-v2-1s2s3s-vs-centrality-wave.pdf}
	%\end{center}
	%\caption{$v_2$ vs centrality.}
	%\label{fig11}
%\end{figure}
%-------------------------

%--------------------------
%\begin{figure}[ht]
	%\begin{center}
	%	\includegraphics[width=0.8\linewidth]{figures/17-raavsnpart-log_aniso-iso-comparison.jpg}
	%\end{center}
%	\caption{Aniso and Iso comparison.}
	%\label{fig12}
%\end{figure}
%-------------------------

%%%%%%%%%%%%%%%%%%%%%%%%%%%%%%%%%%%%%%%%%%%%%%%%%%
\bibliographystyle{JHEP}
\bibliography{anisohqqd}
%%%%%%%%%%%%%%%%%%%%%%%%%%%%%%%%%%%%%%%%%%%%%%%%%%

\end{document}